\documentclass[aps,prl,reprint,superscriptaddress]{revtex4-2}

\usepackage{graphicx}   
\usepackage{amsmath}    
\usepackage{amssymb}
\usepackage{mathrsfs}   
\usepackage{xcolor}
\usepackage{pdfpages}
\usepackage{pgffor}

\makeatletter
\AtBeginDocument{\let\LS@rot\@undefined}
\makeatother

\def\supplementfilename{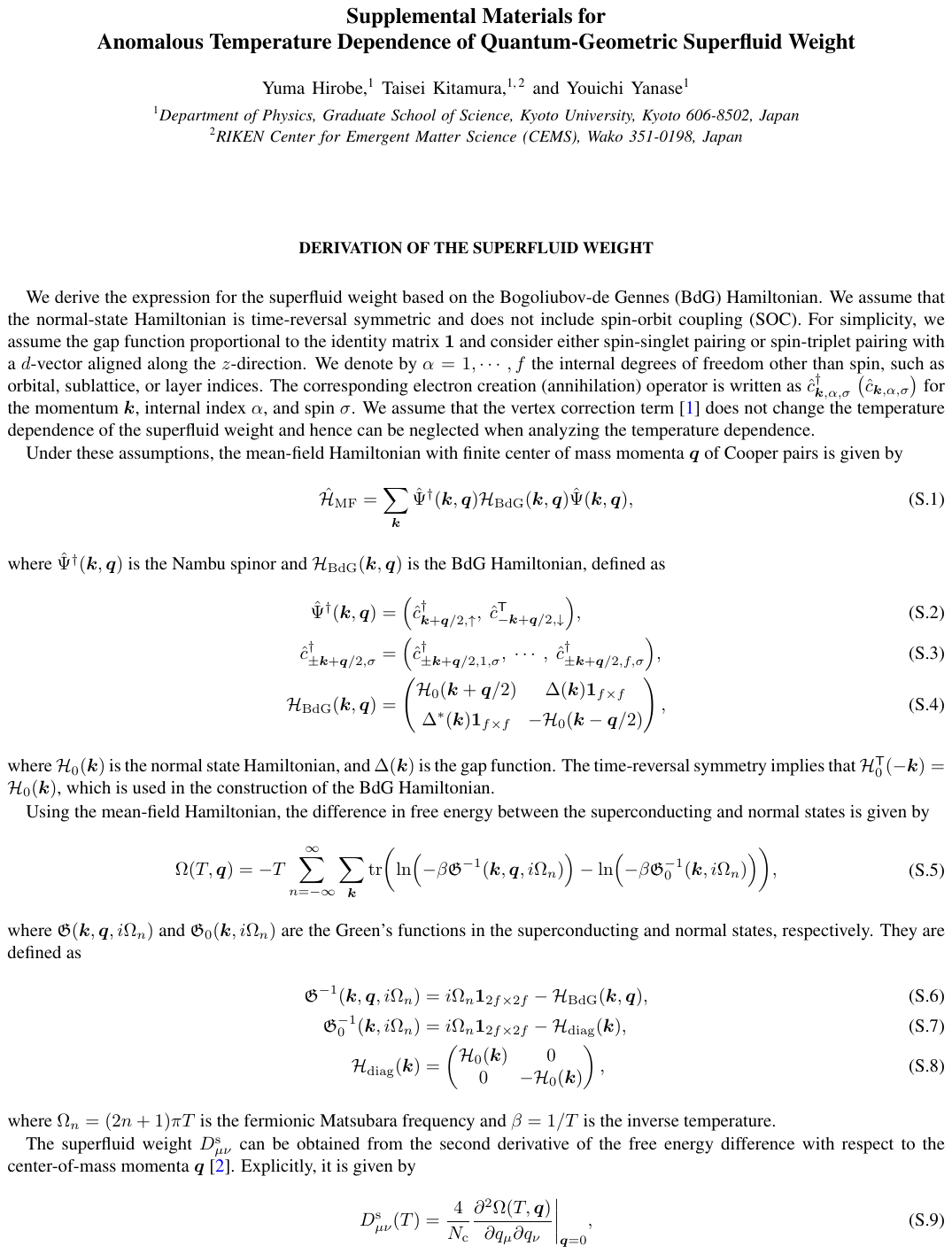}

\pdfximage{\supplementfilename}
\def\numbersupplementpages{\the\pdflastximagepages}

\newif\ifarXiv
\arXivtrue 

\usepackage{braket}
\usepackage{times,multirow,amsfonts,bm,xspace,pifont}
\usepackage{soul}
\usepackage{ulem}
\usepackage{mathtools}

\DeclarePairedDelimiter{\abs}{\lvert}{\rvert}

\renewcommand{\Re}{\operatorname{Re}}

\newcommand{\dd}{\mathrm{d}}

\newcommand{\vk}{\bm{k}}

\newcommand{\ch}{\hat{c}}
\newcommand{\cdh}{\hat{c}^{\dagger}}

\newcommand{\Hch}{\hat{\mathcal{H}}}

\newcommand{\vch}{\hat{\bm{c}}}
\newcommand{\vcdh}{\hat{\bm{c}}^{\dagger}}
\newcommand{\vPsih}{\hat{{\Psi}}}
\newcommand{\vPsidh}{\hat{{\Psi}}^{\dagger}}

\newcommand{\Hc}{\mathcal{H}}

\newcommand*\BdG{{\rm BdG}}

\usepackage{hyperref}

\hypersetup{% hyperref options
 setpagesize=false,%
 colorlinks=true,%
 linkcolor=blue,
 citecolor=blue,
}

\begin{document}

% Title of paper
\title{Anomalous Temperature Dependence of Quantum-Geometric Superfluid Weight}

%authors
\author{Yuma Hirobe}
\email[]{hirobe.yuma.46w@st.kyoto-u.ac.jp}
\affiliation{Department of Physics, Graduate School of Science, Kyoto University, Kyoto 606-8502, Japan}

\author{Taisei Kitamura}
%\email[]{kitamura.taisei.67m@st.kyoto-u.ac.jp}
\affiliation{Department of Physics, Graduate School of Science, Kyoto University, Kyoto 606-8502, Japan}
\affiliation{RIKEN Center for Emergent Matter Science (CEMS), Wako 351-0198, Japan}

\author{Youichi Yanase}
%\email[]{yanase@scphys.kyoto-u.ac.jp}
\affiliation{Department of Physics, Graduate School of Science, Kyoto University, Kyoto 606-8502, Japan}

\date{\today}

\begin{abstract}
The symmetry of Cooper pairs encodes key information about superconductivity and has been widely studied through the temperature dependence of the superfluid weight. However, in systems dominated by quantum geometry, conventional theories miss its essential properties. We study the temperature dependence of the quantum-geometric superfluid weight and classify the relationship to the superconducting symmetry and band structures. The obtained power laws are different from conventional behavior, and unconventional superconductivity in twisted multilayer graphene is discussed. Our findings provide insights into the superconducting symmetry and the pairing mechanism via quantum geometry.
\end{abstract}

\maketitle

\textit{Introduction---}
Determining the symmetry of the order parameter is crucial for studies of quantum condensed matter. The symmetry of superconductivity, which reflects the electronic structure and interactions, is essential for understanding the properties and microscopic mechanism of superconductivity. The measurement of superfluid weight has long been a powerful tool to identify superconducting symmetry~\cite{Gross1986-et,Sigrist1991-ht,Chandrasekhar1993-wb,Prozorov2006-ik}. The temperature dependence of the superfluid weight has been successfully analyzed for various superconductors, demonstrating its effectiveness~\cite{Harshman1989-gf, Harshman1990-re, Annett1991-bd, Hardy1993-sq, Bonalde2000-wg, Kim2002-bv, Carrington2003-hw, Ozcan2003-ow, Bonalde2005-qs, Malone2009-za, Gasparov2009-vq, Khasanov2010-zg, Fernandes2011-ar, Takahashi2011-hc, Kim2011-cx, Kim2012-px, Mazidian2013-ph, Gannon2015-rk, Smylie2016-lq, Smylie2017-il, Kim2018-fo, Biswas2018-rq, Metz2019-vv, Yao2019-oz, Wakamatsu2020-zh, Bae2021-ac, Iguchi2023-lc, Ishihara2023-yd}. In particular, the power-law behavior of the superfluid weight indicates the nodal structure of the superconducting gap, a hallmark of unconventional superconductivity~\cite{Sigrist1991-ht}.

However, conventional theories of superfluid weight can not be applied to systems whose properties are governed by quantum geometry. For example, recent experiments have observed the superfluid weight in twisted bilayer and trilayer graphene~\cite{Tian2023,Tanaka2025-sy, Banerjee2025-ew}, and the measured value for bilayer graphene is much larger than the value predicted based on Fermi liquid theory~\cite{Tian2023,Tanaka2025-sy}. The contradiction can be solved by taking into account the superfluid weight arising from the quantum-geometric properties of Bloch electrons, that is, the quantum-geometric superfluid weight~\cite{Peotta2015-tr,Liang2017-rt,Hu2019-me,Julku2020-ko,Xie2020-oe,Rossi2021,Torma2022}. The quantum-geometric superfluid weight dominates the superfluid responses in flat-band systems, although Fermi liquid theory inadequately predicts vanishing superfluid weight.

In conventional theories of superconductivity, the superfluid weight is primarily determined by the group velocity of electrons near the Fermi surface~\cite{Scalapino1993-wk}, and well-established theories adequately describe the magnitude and temperature dependence in many superconductors. However, in flat-band systems such as twisted multilayer graphene, the group velocity is suppressed, and quantum geometry becomes the dominant origin of the superfluid weight~\cite{Peotta2015-tr,Liang2017-rt}. The quantum-geometric superfluid weight can also give a sizable contribution in some non-flat-band systems~\cite{Kitamura_FeSe}. In these systems, the temperature dependence of the superfluid weight is expected to be different from that predicted by neglecting quantum geometry. Since recent experiments of the superfluid weight provide key information on the symmetry of superconductivity in twisted multilayer graphene, it is desirable to develop theories that clarify the relationship between the quantum-geometric superfluid weight and unconventional superconductivity. Although a specific model has been analyzed~\cite{Bernhard2025-hi}, we lack a general framework linking the order-parameter symmetry to the low-temperature scaling of the quantum-geometric superfluid weight. 

In this Letter, we clarify the missing relationship between the gap structure and the superfluid weight by theoretically analyzing the temperature dependence of the quantum-geometric superfluid weight under various band structures. We uncover scaling laws overlooked in conventional theories and provide a theoretical basis for interpreting recent experimental observations~\cite{Tanaka2025-sy,Banerjee2025-ew}. Our findings will advance studies of the superconducting symmetry and pairing mechanism in twisted multilayer graphene. Moreover, this work reveals comprehensive scaling laws that are applicable beyond flat-band systems and in turn provides a pathway for identifying novel superconducting materials and phenomena via quantum geometry.

\begin{table*}[htbp]
  \centering
  \setlength{\tabcolsep}{10pt}
  \renewcommand{\arraystretch}{1.5}
  \begin{tabular}{c c c c}
  \hline\hline
  gap structure & $\delta D^{\mathrm{conv}}_{\mu\nu}$ & $\delta D^{\mathrm{geom}}_{\mu\nu}$ (flat band) & $\delta D^{\mathrm{geom}}_{\mu\nu}$ (dispersive band) \\
  \hline 
  full gap & $T^{-1/2}e^{-\Delta/T}$ & $e^{-\Delta/T}$ & $T^{1/2} e^{-\Delta/T}$ \\
  point node & $T^{2/l}$ & $T^{2/l + 1}$ & $T^{2/l + 2}$ \\
  line node (w/o crossing) & $T^{1/l}$ & $T^{1/l + 1}$ & $T^{1/l + 2}$ \\
  line node (with crossing) & $-T^{1/l}\ln T$ & $-T^{1/l + 1} \ln T$ & $-T^{1/l + 2} \ln T$ \\
  \hline\hline
  \end{tabular}
  \caption{Scaling laws of the conventional and quantum-geometric contributions to the superfluid weight in the absence of band crossing. For line and point nodes, $l = 1$ corresponds to the linear node, whereas $l = 2$ denotes the quadratic node. Line nodes are classified based on the presence or absence of crossing of multiple line nodes on the Fermi surface. We show the low-temperature behaviors of quantum-geometric superfluid weight for the conventional dispersive bands and for the flat bands. Conventional superfluid weight vanishes for the flat bands.
  }
  \label{tab:temp_dep_no_crossing}
\end{table*}

\textit{Superfluid weight---}
We study the superfluid weight of superconductors using mean-field theory for the intra-band Cooper pairing state. We assume a time-reversal symmetric normal state Hamiltonian without spin-orbit coupling (SOC) and a gap function proportional to the identity matrix \(\bm{1}\), considering either spin-singlet pairing or spin-triplet pairing with a d-vector aligned along the {\it z}-direction. We denote by $\alpha = 1, \cdots, f$ the internal degrees of freedom other than spin, such as orbital, sublattice, or layer indices. The corresponding electron creation (annihilation) operator is written as $\cdh_{\vk,\alpha,\sigma}\ \bigl(\ch_{\vk,\alpha,\sigma}\bigr)$ for the momentum $\vk$, internal index $\alpha$, and spin $\sigma$. 

Under this assumption, the superfluid weight can be calculated from the mean-field Hamiltonian~\cite{Scalapino1993-wk,Taylor2006-ti}:
\begin{equation}
  \Hch_{\mathrm{MF}} = \sum_{\vk}\vPsidh(\vk)\Hc_{\BdG}(\vk)\vPsih(\vk),
\end{equation}
where $\vPsidh(\vk)$ is the Nambu spinor and $\Hc_{\mathrm{BdG}}(\vk)$ is the Bogoliubov-de Gennes (BdG) Hamiltonian, defined as
\begin{align}
  \vPsidh(\vk) &= \Bigl(\cdh_{\vk,\uparrow},\ \ch_{-\vk,\downarrow}^{\mathsf{T}}\Bigr),\\
  \cdh_{\pm\vk,\sigma} &= \Bigl(\cdh_{\pm\vk,1,\sigma},\ \cdots \ ,\ \cdh_{\pm\vk,f,\sigma}\Bigr), 
\end{align}
\begin{equation}
  \Hc_{\BdG}(\vk) = 
  \begin{pmatrix}
    \Hc_0(\vk) & \Delta(\vk)\bm{1}_{f\times f} \\[1ex]
    \Delta^*(\vk)\bm{1}_{f\times f} & -\Hc_0(\vk)
  \end{pmatrix},
  \label{eq:BdG_Hamiltonian}
\end{equation}
with the normal state Hamiltonian \(\Hc_0(\vk)\) and the gap function \(\Delta(\vk)\). The superfluid weight \(D_{\mu\nu}\) can be decomposed into the conventional contribution and the quantum-geometric contribution~\cite{Peotta2015-tr,Liang2017-rt,Julku2018-jr,Huhtinen2022-mu,Kitamura2022-br}:
\begin{equation}
  D_{\mu\nu}(T) = D_{\mu\nu}^{\mathrm{conv}}(T) + D_{\mu\nu}^{\mathrm{geom}}(T).
\end{equation}
These contributions are given by
\begin{align}
  &D_{\mu\nu}^{\mathrm{conv}}(T)=\int_{\mathrm{BZ}}\dfrac{\dd^d \vk}{(2\pi)^d}\sum_{n}  \notag\\
  &\biggl[2f^{\prime}(E_{n}(\vk)) + \frac{f(-E_{n}(\vk))-f(E_{n}(\vk))}{E_{n}(\vk)}\biggr]\\
  &\times\Re\frac{\Delta^{\ast}(\vk)\partial_{\mu}\varepsilon_{n}(\vk)}{E_{n}^2(\vk)}\bigl(\Delta(\vk)\partial_{\nu}\varepsilon_{n}(\vk) - \varepsilon_{n}(\vk)\partial_{\nu}\Delta(\vk)\bigr),\notag
\end{align}
\begin{align}
  &D_{\mu\nu}^{\mathrm{geom}}(T)=\int_{\mathrm{BZ}}\dfrac{\dd^d \vk}{(2\pi)^d}\sum_{n\neq m}\notag\\
  &\ \biggl[\frac{f(-E_{m}(\vk))-f(E_{m}(\vk))}{E_{m}(\vk)} - \frac{f(-E_{n}(\vk))-f(E_{n}(\vk))}{E_{n}(\vk)}\biggr]\notag\\
  &\ \times\frac{\abs{\Delta(\vk)}^2}{\varepsilon_{n}^2(\vk) - \varepsilon_{m}^2(\vk)}\bigl(\varepsilon_{n}(\vk) - \varepsilon_{m}(\vk)\bigr)^2g_{\mu\nu}^{nm}(\vk),
\end{align}
where \(f(E)\) is the Fermi-Dirac distribution function, and \(\varepsilon_{n}(\vk)\) and \(E_{n}(\vk)\) denote the quasiparticle energies in the normal state and the superconducting state, respectively. This decomposition highlights the distinct roles of band dispersion and quantum geometry in determining superfluid weight~\footnote{See Supplemental Material for details of derivation of the superfluid weight and analysis of temperature dependence.}. The quantum-geometric contribution $D_{\mu\nu}^{\mathrm{geom}}(T)$ arises from the band-resolved quantum metric tensor \(g_{\mu\nu}^{nm}(\vk)\) defined as
\begin{equation}
  g_{\mu\nu}^{nm}(\vk) = 2\Re\Braket{u_{n}(\vk)|\partial_{\mu}u_{m}(\vk)}\Braket{\partial_{\nu}u_{m}(\vk)|u_{n}(\vk)},
\end{equation}
by the Bloch wave function $\ket{u_{n}(\vk)}$, an eigenvector of the normal state Hamiltonian $\Hc_0(\vk)$.

\begin{table*}[htbp]
  \centering
  \setlength{\tabcolsep}{10pt} 
  \renewcommand{\arraystretch}{1.5}
  \begin{tabular}{c c c c}
  \hline\hline
  gap structure &  (a) flat \& Dirac bands & (b) Dirac band & (c) dispersive \& dispersive bands \\
  \hline 
  full gap & $e^{-\Delta/T}$ & $e^{-\Delta/T}$ & $T^{-1/2} e^{-\Delta/T}$ \\
  point node & $T$ & $T^{2l - 1}$ & $T^{2/l}$ \\
  line node (w/o crossing ) & $T$ & $T^{2l - 1}$ & $T^{1/l}$ \\
  line node (with crossing) & $T$ & $T^{4l-1}$ & $-T^{1/l}\ln{T}$ \\
  \hline\hline
  \end{tabular}
  \caption{Scaling laws of the quantum-geometric contribution to the superfluid weight in the presence of band crossing at the Fermi level. The parameter $l=1$ and $l=2$ denote the linear and quadratic node, respectively. Band structures (a), (b), and (c) are described in the text and illustrated in Fig.~\ref{fig:bands}. Note that the definition of line and point nodes in the Dirac band [case (b)] follows that in the flat band [case (a)].
  }
  \label{tab:temp_dep_crossing}
\end{table*}

\textit{Scaling law---}
We show the scaling laws of the superfluid weight arising from quantum geometry. We focus on the low temperature region, where the temperature dependence of the gap function can be neglected~\cite{Gross1986-et} and the superfluid weight is affected solely by thermal excitations of Bogoliubov quasiparticles. To obtain the scaling laws, we only need to analyze the reduction in the quantum-geometric contribution due to finite-temperature effects, 
\begin{equation}
  \delta D_{\mu\nu}^{\mathrm{geom}}(T) \coloneqq D_{\mu\nu}^{\mathrm{geom}}(0) - D_{\mu\nu}^{\mathrm{geom}}(T).
\end{equation}
Differentiating this quantity, 
we obtain
\begin{align}
  &\frac{\partial\delta D_{\mu\nu}^{\mathrm{geom}}(T)}{\partial T} = \sum_{m \neq n}\frac{2}{T}\int_{0}^{\infty}\dd E\Biggl[D_{n}(E)\bigl(-f^{\prime}(E)\bigr) \notag \\
  &\times\Braket{\frac{\abs{\Delta(\vk)}^2}{\varepsilon_{m}^2(\vk) - \varepsilon_{n}^2(\vk)}  \bigl(\varepsilon_{m}(\vk) - \varepsilon_{n}(\vk)\bigr)^2  g_{\mu\nu}^{nm}(\vk)}_{E,n}\notag \\
  &\quad + (n\leftrightarrow m)\Biggr],
  \label{eq:tempdep-dos}
\end{align}
where \(D_{n}(E)\) is the density of states of band $n$ at energy \(E\), and \(\braket{\cdots}_{E,n}\) denotes the expectation value for band \(n\) at energy \(E\)~\cite{Bang2009-ho,Lapp2020-jj}, defined as
\begin{align}
  D_{n}(E) &\coloneqq \int_{\mathrm{BZ}}\frac{\dd^d\vk}{(2\pi)^d}\delta\bigl(E - E_{n}(\vk)\bigr),\\
  \braket{\mathcal{O}(\vk)}_{E,n} &\coloneqq \frac{1}{D_{n}(E)}\int_{\mathrm{BZ}}\frac{\dd^{d}\vk}{(2\pi)^{d}}\delta\bigl(E - E_{n}(\vk)\bigr)\mathcal{O}(\vk).
\end{align}
In the special two-band models where \(E_{n}(\vk) = E_{m}(\vk)\) for all $\vk$, Eq.~\eqref{eq:tempdep-dos} is simplified to
\begin{equation}
  \begin{split}
    &\frac{\partial\delta D_{\mu\nu}^{\mathrm{geom}}(T)}{\partial T} = \frac{1}{T}\int_{0}^{\infty}\dd E D_{n}(E)f^{\prime\prime}(E)\\
    &\qquad\quad\times\Braket{\frac{\abs{\Delta(\vk)}^2}{E_n(\vk)}\bigl(\varepsilon_{n}(\vk) - \varepsilon_{m}(\vk)\bigr)^2g_{\mu\nu}^{nm}(\vk)}_{E,n}.
  \end{split}
  \label{eq:tempdep-dos2}
\end{equation}
We use this equation later for the calculations of cases (b) and (c) in Table~\ref{tab:temp_dep_crossing}. Hereafter, we assume that \(\bigl(\varepsilon_{m}(\vk) - \varepsilon_{n}(\vk)\bigr)^2g_{\mu\nu}^{nm}(\vk)\) remains finite at the nodes of the excitation gap.
 
Then, by analyzing Eqs.~\eqref{eq:tempdep-dos} and~\eqref{eq:tempdep-dos2} for $\delta D_{\mu\nu}^{\rm geom}$, we show the scaling law of quantum-geometric superfluid weight. Since Eq.~\eqref{eq:tempdep-dos} is the Fermi surface term, we only need to consider the bands near the Fermi level. We separately analyze the following two cases because the scaling law depends on the band structure. 

1. Absence of band crossing at the Fermi level: 
  When no band crossing point exists at the Fermi level, we can separately evaluate the contribution of band \(n\) that crosses the Fermi level. When the band \(m\) is well separated from the Fermi level, like the conventional contribution, the temperature dependence of the quantum-geometric contribution from the band $m$ is suppressed by the factor \(e^{-W_{m}/T}\) with
  \begin{equation}
    W_{m} \coloneq \min_{\vk_{\mathrm{F}}}\bigl(\sqrt{\varepsilon_{m}^2(\vk) + \abs{\Delta(\vk)}^2} - \abs{\Delta(\vk)}\bigr),
  \end{equation}
  and is negligible at low temperatures \(T \ll W_{m}\). Then, \(D_{n}(E)\) and \(\braket{\cdots}_{E,n}\) in the low-energy region need to be evaluated to obtain the low-temperature behavior. Consequently, we find that \(\delta D_{\mu\nu}^{\mathrm{geom}}(T)\) follows the temperature dependence listed in Table~\ref{tab:temp_dep_no_crossing}, where we show the scaling laws for the conventional dispersive band and for the flat band. The latter may be relevant to the magic-angle twisted bilayer graphene (MATBG), which hosts an isolated flat band~\cite{Bistritzer2011-bi,Cao2018-qu,Cao2018-gn}.

  The scaling laws are classified by the gap structure, that is, the full gap, line nodal gap, and point nodal gap. Here, we define line and point nodes by dimension of nodal region of excitations: When the dimension of nodal region is reduced by one (two) from that of Fermi surface, we denote the line (point) node. In two-dimensional flat band systems, the Fermi surface is the whole two-dimensional Brillouin zone, and the line (point) node corresponds to the nodal structure of the gap function \(|\Delta(\vk)|\). In three-dimensional dispersive bands, the definition coincides with the nodal structure of Bogoliubov quasiparticles $E_n(\vk)$. In two-dimensional dispersive bands, the excitation gap closes at a point in the case of the {\it line node}. In Table~\ref{tab:temp_dep_no_crossing}, the line nodal gap is furthermore distinguished depending on the presence or absence of crossing of multiple nodal lines. We see that the logarithmic correction \(-\ln T\) appears in the case of crossing line nodes. The logarithmic factor leads to reduction in the power of temperature dependence~\cite{Mazidian2013-ph}, when we perform the power-law fitting as is often done in the analysis of experimental data.

\begin{figure}[htbp]
  \centering
  \includegraphics[width=0.48\textwidth]{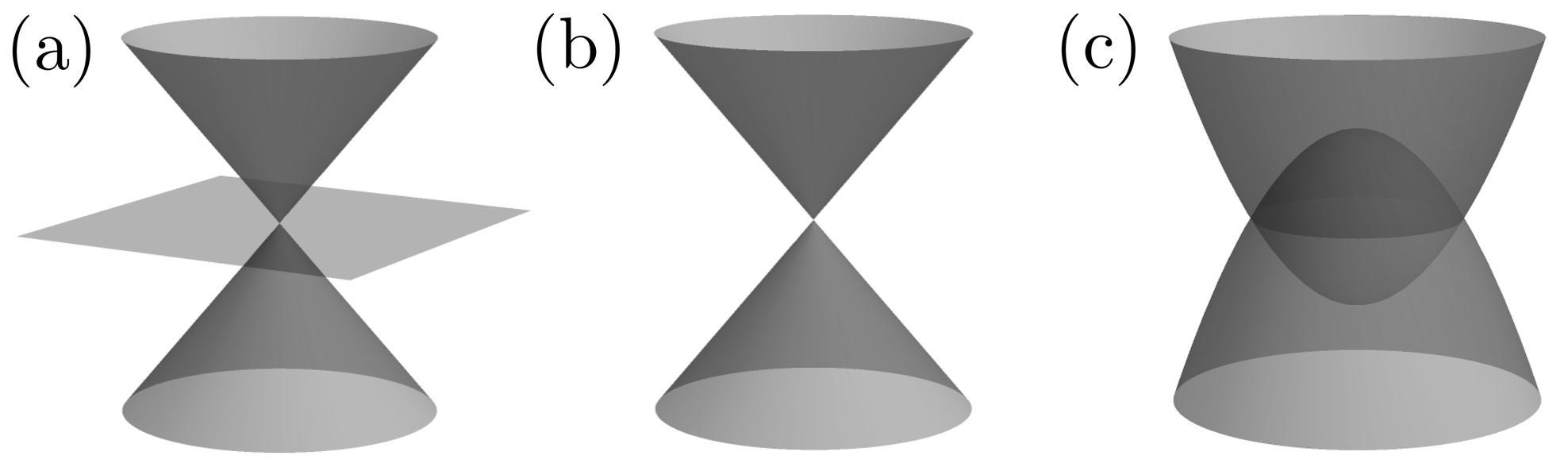}
  \caption{Representative band structures with band crossing. (a) A flat band crosses a Dirac band. (b) A Dirac band formed by crossing linear bands. (c) Two dispersive bands cross each other.
  }
  \label{fig:bands}
\end{figure}

2. Presence of band crossing at the Fermi level: In  topological (semi)metals such as the Dirac and Weyl electron systems, the band crossing occurs near the Fermi level. When the band crossing point exists at the Fermi level, we have to carefully analyze the multiband effects. We compute the temperature dependence of the superfluid weight in the following cases (Fig.~\ref{fig:bands}): (a) A flat band crosses a Dirac band~\cite{Julku2016-tj,Park2021-bz}. (b) A Dirac band where two linear bands intersect~\cite{Wallace1947-tp,Kopnin2008-jg, He2016-fe}. (c) Two dispersive bands intersect~\cite{Burkov2011-ru, Wu2018-wv, Gao2020-ef, Li2020-kp, Lygouras2025-cs}. The case (a) may be relevant to the magic angle twisted trilayer graphene (MATTG), where the flat band intersects a Dirac band at the Fermi level~\cite{Park2021-bz, Kim2022-wh}.

The obtained results are summarized in Table~\ref{tab:temp_dep_crossing}, which describes the scaling laws when the band crossing point coincides with the nodal point in the gap function. If the band crossing point and the nodal point are separated in the momentum space, the temperature dependence of superfluid weight obeys the scaling laws in Table~\ref{tab:temp_dep_no_crossing}.

\begin{figure}[htbp]
  \centering
  \includegraphics[width=0.48\textwidth]{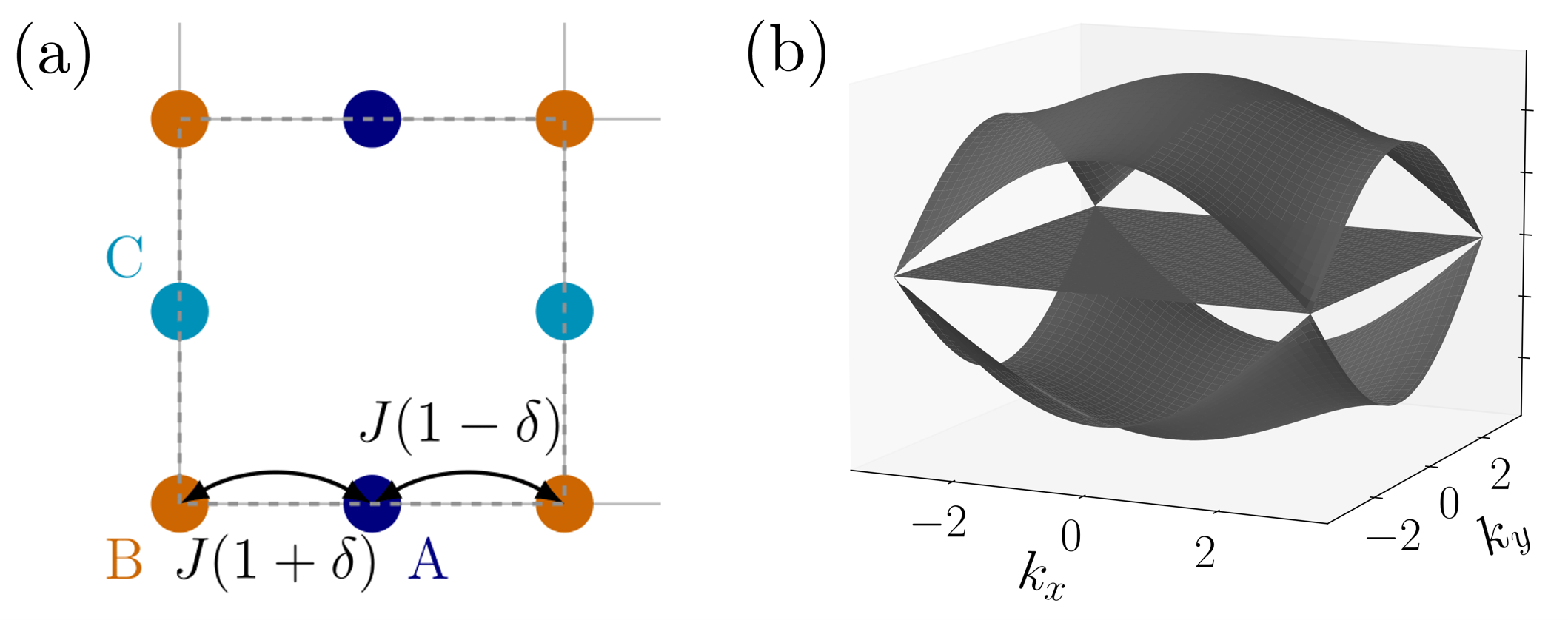}
  \caption{(a) Lattice structure and hopping integrals of the modified Lieb lattice. A unit cell is indicated by the dotted square. (b) Band structure for $\delta=0$, where a flat band intersects a Dirac band at the M point.
  }
  \label{fig:lieb}
\end{figure}

\textit{Application to superconductivity in the Lieb lattice---}
To verify the scaling laws, we evaluate the superfluid weight in a modified Lieb lattice [Fig.~\ref{fig:lieb}(a)]. Since the staggered hopping term can modify the flat band structure in the canonical Lieb lattice~\cite{Lieb1989-we, Julku2016-tj, Huhtinen2022-mu, Penttila2025-my} and lead to various band structures, we can numerically examine the scaling laws shown above.

The modified Lieb lattice, illustrated in Fig.~\ref{fig:lieb}(a), consists of three sites per unit cell, labeled A, B, and C. In addition to the nearest-neighbor inter-sublattice hopping \(J\), the staggered hopping \(\delta J\) controls the band structure. The normal state Hamiltonian is given by
\begin{equation}
  \begin{split}
    &\Hch_{\mathrm{Lieb}} = \sum_{\vk\sigma}\vcdh_{\vk\sigma}\Hc_{\mathrm{Lieb}}(\vk)\vch_{\vk\sigma}\\
    &\ = \sum_{\vk\sigma}(\cdh_{\vk A\sigma},\cdh_{\vk B\sigma},\cdh_{\vk C\sigma})
    \begin{pmatrix}
      0 & a_{\vk} & 0\\
      a_{\vk}^{\ast} & 0 & b_{\vk}\\
      0 & b_{\vk}^{\ast} & 0
    \end{pmatrix}
    \begin{pmatrix}
      \ch_{\vk A\sigma}\\
      \ch_{\vk B\sigma}\\
      \ch_{\vk C\sigma}
    \end{pmatrix},
  \end{split}
\end{equation}
where
\begin{align}
  a_{\vk} &= 2J(\cos{k_x/2} + i\delta\sin{k_x/2}),\\
  b_{\vk} &= 2J(\cos{k_y/2} + i\delta\sin{k_y/2}).
\end{align}
We set the unit of energy as $J=1$. In the canonical Lieb lattice, \(\delta=0\), the flat band crosses a Dirac point at the M point [Fig.~\ref{fig:lieb}(b)], realizing the case (a) in Table~\ref{tab:temp_dep_crossing}. When \(\delta \neq 0\), the band crossing is removed while the flat band remains.

We consider the following two gap functions:
\begin{align}
  \Delta^{(1)}(\vk) &= \Delta_0(\cos{k_x} - \cos{k_y}),\\
  \Delta^{(2)}(\vk) &= \Delta_0(\cos{k_x} + \cos{k_y} - 1).
\end{align}
The gap function of $d$-wave superconductivity \(\Delta^{(1)}(\vk)\) has line nodes that cross each other at the \(\Gamma\) and M points, whereas \(\Delta^{(2)}(\vk)\) for extended $s$-wave superconductivity has a single line node. Modeling the temperature dependence of the gap function as
\begin{equation}
  \Delta_0(T) = \Delta_0(0)\tanh\Biggl(\frac{\pi T_{\mathrm{c}}}{\Delta_{0}(0)}\sqrt{a\biggl(\frac{T_{\mathrm{c}}}{T} - 1\biggr)}\Biggr), 
\end{equation}
with $\Delta_{0}(0) = 2.46T_{\mathrm{c}}$ and $a = 1.56$~\cite{Gross1986-et}, we compute the temperature dependence of the superfluid weight.

\begin{figure}[htbp]
  \centering 
\includegraphics[width=0.475\textwidth]{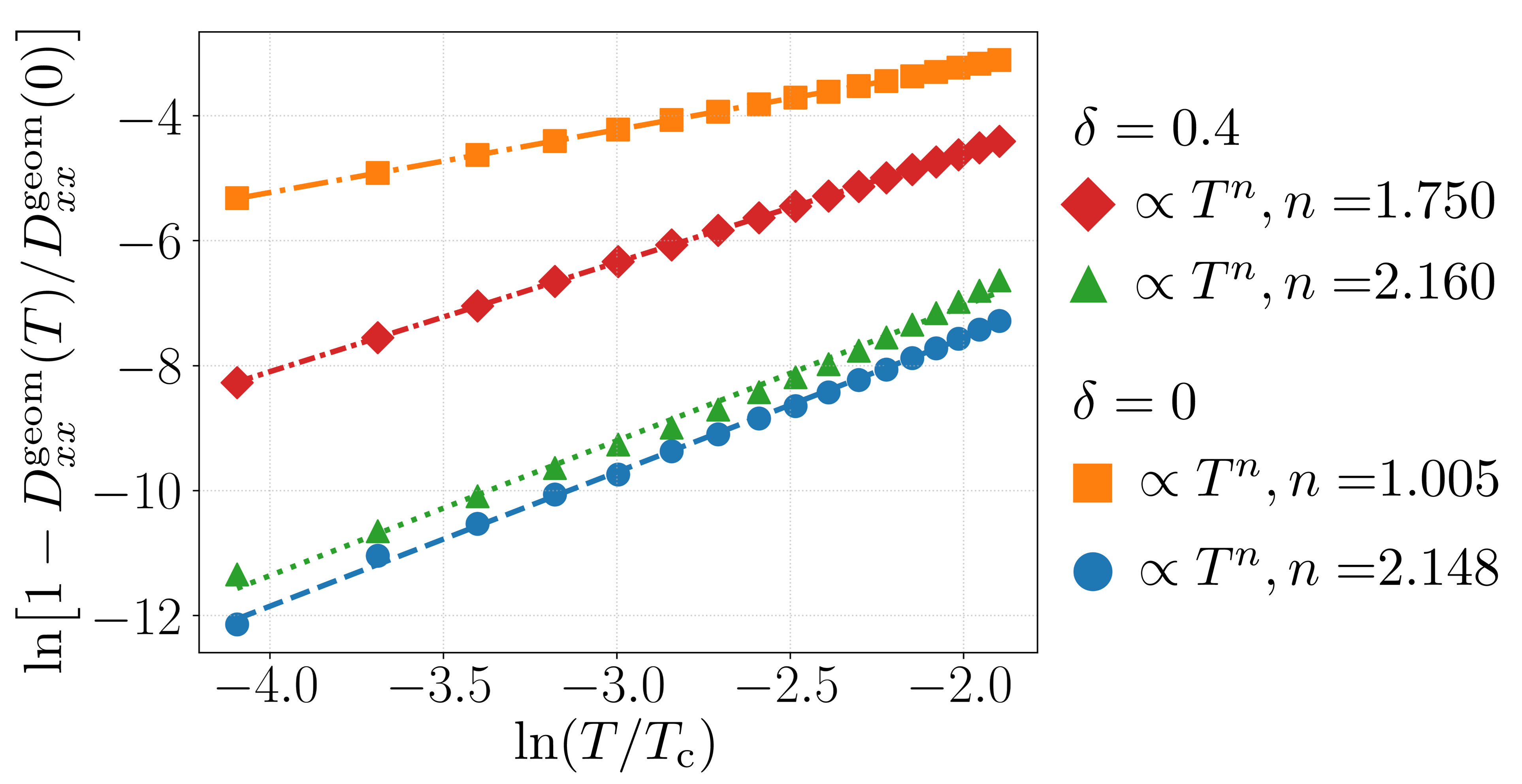}
  \caption{Temperature dependence of the superfluid weight in the modified Lieb lattice. The band crossing is absent for $\delta=0.4$, while it is present for $\delta = 0$. For the $d$-wave gap function \(\Delta^{(1)}(\vk)\), diamonds and squares show the results for $\delta=0.4$ and $0$, respectively. For the extended $s$-wave gap function \(\Delta^{(2)}(\vk)\), triangles and circles show the results for $\delta=0.4$ and $0$, respectively. We set $\Delta_{0}(0)=0.4$. Lines are the results of power-law fitting.
  }
  \label{fig:lieb1}
  \hfill
\end{figure}

First, to study the system without band crossing at the Fermi level, we set \(\delta =  0.4\), where the isolated flat band appears. The results of the superfluid weight are shown in Fig.~\ref{fig:lieb1}. We see that the superfluid weight follows \(T^{1.75}\) for the gap function \(\Delta^{(1)}(\vk)\) and \(T^{2.16}\) for \(\Delta^{(2)}(\vk)\), consistent with the predicted scaling laws of \(-T^2\ln{T}\) and \(T^2\), respectively (see Table~\ref{tab:temp_dep_no_crossing}). 

Next, we study the system with band crossing at the Fermi level by setting $\delta=0$, that is, the canonical Lieb lattice [Fig.~\ref{fig:lieb}(b)]. The gap function \(\Delta^{(1)}(\vk)\) realizes the case (a) in Table~\ref{tab:temp_dep_crossing}. Consistent with Table~\ref{tab:temp_dep_crossing}, the superfluid weight shows \(T\)-linear behavior, whereas it follows \(T^{2.148}\) for the gap function \(\Delta^{(2)}(\vk)\) in agreement with the theoretical prediction of \(T^2\)-scaling for the flat band system in Table~\ref{tab:temp_dep_no_crossing}. 

As shown above, the numerical results in the modified Lieb lattice confirm the classification of the scaling laws for the quantum-geometric superfluid weight, demonstrating the consistency between analytical and numerical approaches.

\textit{Discussion---}
Recently, the temperature dependence of the superfluid weight has been measured in the MATBG and MATTG~\cite{Banerjee2025-ew,Tanaka2025-sy}. In MATBG, an isolated flat band exists at the Fermi level~\cite{Bistritzer2011-bi,Cao2018-qu,Cao2018-gn}, whereas in MATTG, the flat band intersects a Dirac band at the Fermi level~\cite{Park2021-bz, Kim2022-wh}. Since both systems host flat bands, the quantum-geometric contribution is expected to play a significant role in the superfluid weight. 

Experiments have reported a \(T^2\)-dependence in MATBG~\cite{Tanaka2025-sy} whereas a \(T\)-linear dependence in MATTG~\cite{Banerjee2025-ew}. By comparison of these experiments with the theoretical results of this study, we can infer the gap structure in the superconducting states of twisted graphene systems. Our scaling laws for an isolated flat band system in Table~\ref{tab:temp_dep_no_crossing} can apply to MATBG, and then a linear line node and a quadratic point node are consistent with the experiment. In MATTG, there is a flat band along with a Dirac point at the $K$ point, and thus case (a) in Table~\ref{tab:temp_dep_crossing} implies the presence of the gap node at the $K$ point.

If we assume that the same superconducting gap structure is realized in MATBG and MATTG, the gap structure is expected to be either a linear line node or a quadratic point node passing through the $K$ point. This constraint limits the possible symmetry of superconductivity. Taking into account the \(D_{3}\) point group symmetry of MATBG and MATTG~\cite{Sharpe2019-fi, Long2023-iq, Yu2024-gf}, gap functions that produce linear line nodes passing through the $K$ point include the nematic $d$-wave and $p$-wave states, such as with $d_{x^2-y^2}$, $d_{xy}$, $p_x$ and $p_y$ symmetry. A quadratic point node at the $K$ point is consistent with the chiral $d$-wave state with $d_{x^2-y^2} + i d_{xy}$ symmetry~\cite{Black-Schaffer2014-di}.

Furthermore, the observation of twofold anisotropy in the superconducting phase of MATBG~\cite{Cao2021-ai} suggests that the superconducting order parameter is non-chiral. The combination of all implications points to a nematic $d$-wave or $p$-wave state. Recent experimental studies also support nodal superconductivity in MATBG and MATTG~\cite{Oh2021-ej,Kim2022-wh,Park2025-zv}. These discussions demonstrate that the superfluid weight can serve as a probe of the symmetry of superconductivity in twisted multilayer graphene and other flat band systems by properly accounting for quantum geometry. 

\textit{Conclusion---}
We have investigated how the superfluid weight arising from the quantum geometry depends on the temperature in the context of unconventional superconductivity. Our results reveal that the quantum-geometric superfluid weight exhibits characteristic low-temperature behaviors depending on the nodal gap structure. The numerical calculations have verified the analytically derived power laws. The temperature scaling is different from the conventional superfluid weight, highlighting a manifestation of quantum geometry in the superconducting responses. 

Our results provide a framework for interpreting recent experimental studies in MATBG and MATTG, and the symmetry of superconductivity is concluded to be unconventional $d$-wave or $p$-wave. This study not only contributes to a deeper understanding of the superconducting states in twisted multilayer graphene but also establishes a theoretical framework applicable to a broader class of materials with significant quantum geometry. 

\begin{acknowledgments}
The authors are grateful to K. Hara, T. Matsumoto, H. Tanaka, K. Shinada, R. Sano, T. Matsushita and M. Tanaka for fruitful discussions.
This work was supported by JSPS KAKENHI (Grant Numbers JP22H01181, JP22H04933, JP23K17353, JP23K22452, JP24K21530, JP24H00007, JP25H01249).
Y.~H. was supported by Iwadare Scholarship Foundation.
\end{acknowledgments}

\bibliography{TempDep}

\ifarXiv
  

\section*{Supplementary material}

# Supplemental Materials for
# Anomalous Temperature Dependence of Quantum-Geometric Superfluid Weight

Yuma Hirobe,[1] Taisei Kitamura,[1,2] and Youichi Yanase[1]

[1]*Department of Physics, Graduate School of Science, Kyoto University, Kyoto 606-8502, Japan*
[2]*RIKEN Center for Emergent Matter Science (CEMS), Wako 351-0198, Japan*

### DERIVATION OF THE SUPERFLUID WEIGHT

We derive the expression for the superfluid weight based on the Bogoliubov-de Gennes (BdG) Hamiltonian. We assume that the normal-state Hamiltonian is time-reversal symmetric and does not include spin-orbit coupling (SOC). For simplicity, we assume the gap function proportional to the identity matrix $\mathbf{1}$ and consider either spin-singlet pairing or spin-triplet pairing with a $d$-vector aligned along the $z$-direction. We denote by $\alpha = 1, \cdots, f$ the internal degrees of freedom other than spin, such as orbital, sublattice, or layer indices. The corresponding electron creation (annihilation) operator is written as $\hat{c}^\dagger_{\bm{k},\alpha,\sigma}$ ($\hat{c}_{\bm{k},\alpha,\sigma}$) for the momentum $\bm{k}$, internal index $\alpha$, and spin $\sigma$. We assume that the vertex correction term [1] does not change the temperature dependence of the superfluid weight and hence can be neglected when analyzing the temperature dependence.

Under these assumptions, the mean-field Hamiltonian with finite center of mass momenta $\bm{q}$ of Cooper pairs is given by

$$\hat{\mathcal{H}}_{\mathrm{MF}} = \sum_{\bm{k}} \hat{\Psi}^\dagger(\bm{k},\bm{q}) \mathcal{H}_{\mathrm{BdG}}(\bm{k},\bm{q}) \hat{\Psi}(\bm{k},\bm{q}), \tag{S.1}$$

where $\hat{\Psi}^\dagger(\bm{k},\bm{q})$ is the Nambu spinor and $\mathcal{H}_{\mathrm{BdG}}(\bm{k},\bm{q})$ is the BdG Hamiltonian, defined as

$$\hat{\Psi}^\dagger(\bm{k},\bm{q}) = \left( \hat{c}^\dagger_{\bm{k}+\bm{q}/2,\uparrow},\ \hat{c}^{\mathsf{T}}_{-\bm{k}+\bm{q}/2,\downarrow} \right), \tag{S.2}$$

$$\hat{c}^\dagger_{\pm\bm{k}+\bm{q}/2,\sigma} = \left( \hat{c}^\dagger_{\pm\bm{k}+\bm{q}/2,1,\sigma},\ \cdots,\ \hat{c}^\dagger_{\pm\bm{k}+\bm{q}/2,f,\sigma} \right), \tag{S.3}$$

$$\mathcal{H}_{\mathrm{BdG}}(\bm{k},\bm{q}) = \begin{pmatrix} \mathcal{H}_0(\bm{k}+\bm{q}/2) & \Delta(\bm{k})\mathbf{1}_{f \times f} \\ \Delta^*(\bm{k})\mathbf{1}_{f \times f} & -\mathcal{H}_0(\bm{k}-\bm{q}/2) \end{pmatrix}, \tag{S.4}$$

where $\mathcal{H}_0(\bm{k})$ is the normal state Hamiltonian, and $\Delta(\bm{k})$ is the gap function. The time-reversal symmetry implies that $\mathcal{H}_0^{\mathsf{T}}(-\bm{k}) = \mathcal{H}_0(\bm{k})$, which is used in the construction of the BdG Hamiltonian.

Using the mean-field Hamiltonian, the difference in free energy between the superconducting and normal states is given by

$$\Omega(T,\bm{q}) = -T \sum_{n=-\infty}^{\infty} \sum_{\bm{k}} \mathrm{tr}\left( \ln\left(-\beta \mathfrak{G}^{-1}(\bm{k},\bm{q},i\Omega_n)\right) - \ln\left(-\beta \mathfrak{G}_0^{-1}(\bm{k},i\Omega_n)\right) \right), \tag{S.5}$$

where $\mathfrak{G}(\bm{k},\bm{q},i\Omega_n)$ and $\mathfrak{G}_0(\bm{k},i\Omega_n)$ are the Green's functions in the superconducting and normal states, respectively. They are defined as

$$\mathfrak{G}^{-1}(\bm{k},\bm{q},i\Omega_n) = i\Omega_n \mathbf{1}_{2f \times 2f} - \mathcal{H}_{\mathrm{BdG}}(\bm{k},\bm{q}), \tag{S.6}$$

$$\mathfrak{G}_0^{-1}(\bm{k},i\Omega_n) = i\Omega_n \mathbf{1}_{2f \times 2f} - \mathcal{H}_{\mathrm{diag}}(\bm{k}), \tag{S.7}$$

$$\mathcal{H}_{\mathrm{diag}}(\bm{k}) = \begin{pmatrix} \mathcal{H}_0(\bm{k}) & 0 \\ 0 & -\mathcal{H}_0(\bm{k}) \end{pmatrix}, \tag{S.8}$$

where $\Omega_n = (2n+1)\pi T$ is the fermionic Matsubara frequency and $\beta = 1/T$ is the inverse temperature.

The superfluid weight $D^{\mathrm{s}}_{\mu\nu}$ can be obtained from the second derivative of the free energy difference with respect to the center-of-mass momenta $\bm{q}$ [2]. Explicitly, it is given by

$$D^{\mathrm{s}}_{\mu\nu}(T) = \frac{4}{N_{\mathrm{c}}} \frac{\partial^2 \Omega(T,\bm{q})}{\partial q_\mu \partial q_\nu} \bigg|_{\bm{q}=0}, \tag{S.9}$$

where $N_c$ is the number of unit cells in the system. Evaluating the term, we find

$$\frac{4}{N_c}\frac{\partial^2 \Omega(T,\boldsymbol{q})}{\partial q_\mu \partial q_\nu}\bigg|_{\boldsymbol{q}=0} = \frac{1}{2N_c}\sum_{\boldsymbol{k}}\sum_{ij=1}^{2f}\frac{f(E_j)-f(E_i)}{E_i-E_j}\Big(\langle\psi_i|\partial_\mu\mathcal{H}_{\text{diag}}|\psi_j\rangle\langle\psi_j|\partial_\nu\mathcal{H}_{\text{diag}}|\psi_i\rangle \qquad (S.10)$$
$$+\langle\psi_i|\partial_\mu\boldsymbol{\Delta}|\psi_j\rangle\langle\psi_j|\partial_\nu\mathcal{H}_{\text{diag}}|\psi_i\rangle - \langle\psi_i|\partial_\mu\mathcal{H}_{\text{diag}}\gamma^z|\psi_j\rangle\langle\psi_j|\partial_\nu\mathcal{H}_{\text{diag}}\gamma^z|\psi_i\rangle\Big),$$

where $E_i(\boldsymbol{k})$ and $|\psi_i(\boldsymbol{k})\rangle$ are the eigenvalues and eigenvectors of the BdG Hamiltonian $\mathcal{H}_{\text{BdG}}(\boldsymbol{k})=\mathcal{H}_{\text{BdG}}(\boldsymbol{k},\boldsymbol{q}=0)$, satisfying

$$\mathcal{H}_{\text{BdG}}(\boldsymbol{k})|\psi_i(\boldsymbol{k})\rangle = E_i(\boldsymbol{k})|\psi_i(\boldsymbol{k})\rangle. \qquad (S.11)$$

The matrices $\boldsymbol{\Delta}(\boldsymbol{k})$ and $\gamma^z$ are defined by

$$\boldsymbol{\Delta}(\boldsymbol{k}) = \begin{pmatrix} 0 & \Delta(\boldsymbol{k})\mathbf{1}_{f\times f} \\ \Delta^*(\boldsymbol{k})\mathbf{1}_{f\times f} & 0 \end{pmatrix}, \qquad (S.12)$$

$$\gamma^z = \sigma^z \otimes \mathbf{1}_{f\times f} = \begin{pmatrix} \mathbf{1}_{f\times f} & 0 \\ 0 & -\mathbf{1}_{f\times f} \end{pmatrix}. \qquad (S.13)$$

We now define the eigenvalues and eigenvectors of the normal state Hamiltonian $\mathcal{H}_0(\boldsymbol{k})$ and the Pauli matrix $\sigma^z$ as follows:

$$\mathcal{H}_0(\boldsymbol{k})|u_n(\boldsymbol{k})\rangle = \varepsilon_n(\boldsymbol{k})|u_n(\boldsymbol{k})\rangle, \qquad (S.14)$$
$$\sigma^z|\pm\rangle = \pm|\pm\rangle. \qquad (S.15)$$

Expanding the eigenstates $|\psi_i(\boldsymbol{k})\rangle$ of the BdG Hamiltonian in terms of the tensor product basis of $|u_n(\boldsymbol{k})\rangle$ and $|\pm\rangle$, we write

$$|\psi_i(\boldsymbol{k})\rangle = \sum_{n=1}^{f}\Big(w_{in}^p|+\rangle \otimes |u_n(\boldsymbol{k})\rangle + w_{in}^h|-\rangle \otimes |u_n(\boldsymbol{k})\rangle\Big). \qquad (S.16)$$

Using this expansion, the superfluid weight can be decomposed into three contributions as

$$D_{\mu\nu}^{\text{s}} = D_{\mu\nu}^{\text{conv}} + D_{\mu\nu}^{\text{geom}}, \qquad (S.17)$$

where $D_{\mu\nu}^{\text{conv}}$ and $D_{\mu\nu}^{\text{geom}}$ represent, respectively, the contributions from the band dispersion and quantum geometry. To evaluate these contributions, we introduce the following quantities:

$$W_{nm}^{ls} = \sum_{ij=1}^{2f}\frac{f(E_j)-f(E_i)}{E_i-E_j}w_{in}^{p*}w_{jm}^{p}w_{jl}^{h*}w_{is}^{h}, \qquad (S.18)$$

$$V_{nm}^{ls,p} = \sum_{ij=1}^{2f}\frac{f(E_j)-f(E_i)}{E_i-E_j}w_{in}^{p*}w_{jm}^{p}w_{jl}^{p*}w_{is}^{h}, \qquad (S.19)$$

$$V_{nm}^{ls,h} = \sum_{ij=1}^{2f}\frac{f(E_j)-f(E_i)}{E_i-E_j}w_{in}^{p*}w_{jm}^{p}w_{jl}^{h*}w_{is}^{p}, \qquad (S.20)$$

$$S_{nm}^{ls,p} = \sum_{ij=1}^{2f}\frac{f(E_j)-f(E_i)}{E_i-E_j}w_{in}^{h*}w_{jm}^{h}w_{jl}^{p*}w_{is}^{h}, \qquad (S.21)$$

$$S_{nm}^{ls,h} = \sum_{ij=1}^{2f}\frac{f(E_j)-f(E_i)}{E_i-E_j}w_{in}^{h*}w_{jm}^{h}w_{jl}^{h*}w_{is}^{p}, \qquad (S.22)$$

$$\left[j_\mu^p(\boldsymbol{k})\right]_{nm} = -\left[j_\mu^h(-\boldsymbol{k})\right]_{nm} = \langle u_n|\partial_\mu\mathcal{H}_0(\boldsymbol{k})|u_m\rangle = \partial_\mu\varepsilon_n(\boldsymbol{k})\delta_{nm} + \big(\varepsilon_m(\boldsymbol{k})-\varepsilon_n(\boldsymbol{k})\big)\langle u_n|\partial_\mu u_m\rangle, \qquad (S.23)$$

$$\left[\delta\Delta_\mu(\boldsymbol{k})\right]_{nm} = \langle u_n|\partial_\mu\boldsymbol{\Delta}(\boldsymbol{k})|u_m\rangle = \partial_\mu\Delta(\boldsymbol{k})\delta_{nm}, \qquad (S.24)$$

$$\left[\delta\Delta_\mu^\dagger(\boldsymbol{k})\right]_{nm} = \langle u_n|\partial_\mu\boldsymbol{\Delta}^\dagger(\boldsymbol{k})|u_m\rangle = \partial_\mu\Delta^*(\boldsymbol{k})\delta_{nm}. \qquad (S.25)$$

Using the definitions above, the conventional contribution $D_{\mu\nu}^{\text{conv}}$ can be decomposed into two parts:

$$D_{\mu\nu}^{\text{conv}} = D_{\mu\nu}^{\text{conv1}} + D_{\mu\nu}^{\text{gap1}}, \tag{S.26}$$

$$D_{\mu\nu}^{\text{conv1}} = \frac{2}{N_c} \sum_{\bm{k}} \sum_{nm} W_{nn}^{mm} \left[j_\mu^p(\bm{k})\right]_{nn} \left[j_\nu^h(-\bm{k})\right]_{mm}, \tag{S.27}$$

$$\begin{aligned} D_{\mu\nu}^{\text{gap1}} = \frac{1}{2N_c} \sum_{\bm{k}} \sum_{nls} \Big( & V_{nn}^{ls,p} \left[j_\mu^p(\bm{k})\right]_{nn} \left[\delta\Delta_\nu(\bm{k})\right]_{ls} + V_{nn}^{ls,h} \left[j_\mu^p(\bm{k})\right]_{nn} \left[\delta\Delta_\nu^\dagger(\bm{k})\right]_{ls} \\ & + S_{nn}^{ls,p} \left[j_\mu^h(-\bm{k})\right]_{nn} \left[\delta\Delta_\nu(\bm{k})\right]_{ls} + S_{nn}^{ls,h} \left[j_\mu^h(-\bm{k})\right]_{nn} \left[\delta\Delta_\nu^\dagger(\bm{k})\right]_{ls} \Big), \end{aligned} \tag{S.28}$$

where $D_{\mu\nu}^{\text{conv1}}$ represents the contribution of the group velocity and $D_{\mu\nu}^{\text{gap1}}$ represents the modification of the Fermi-liquid contribution due to the gap function. The quantum-geometric contribution $D_{\mu\nu}^{\text{geom}}$ is further decomposed into four parts as follows.

$$D_{\mu\nu}^{\text{geom}} = D_{\mu\nu}^{\text{geom1}} + D_{\mu\nu}^{\text{geom2}} + D_{\mu\nu}^{\text{multi}} + D_{\mu\nu}^{\text{gap2}}, \tag{S.29}$$

$$D_{\mu\nu}^{\text{geom1}} = \frac{2}{N_c} \sum_{\bm{k}} \sum_{n \neq m} W_{nm}^{mn} \left[j_\mu^p(\bm{k})\right]_{nm} \left[j_\nu^h(-\bm{k})\right]_{mn}, \tag{S.30}$$

$$D_{\mu\nu}^{\text{geom2}} = \frac{2}{N_c} \sum_{\bm{k}} \sum_{n \neq m, l \neq s, n \neq s, m \neq l} W_{nm}^{ls} \left[j_\mu^p(\bm{k})\right]_{nm} \left[j_\nu^h(-\bm{k})\right]_{ls}, \tag{S.31}$$

$$D_{\mu\nu}^{\text{multi}} = \frac{2}{N_c} \sum_{\bm{k}} \sum_{n, l \neq s} \Big( W_{nn}^{ls} \left[j_\mu^p(\bm{k})\right]_{nn} \left[j_\nu^h(-\bm{k})\right]_{ls} + W_{ls}^{nn} \left[j_\mu^p(\bm{k})\right]_{ls} \left[j_\nu^h(-\bm{k})\right]_{nn} \Big), \tag{S.32}$$

$$\begin{aligned} D_{\mu\nu}^{\text{gap2}} = \frac{1}{2N_c} \sum_{\bm{k}} \sum_{n \neq m, ls} \Big( & V_{nm}^{ls,p} \left[j_\mu^p(\bm{k})\right]_{nm} \left[\delta\Delta_\nu(\bm{k})\right]_{ls} + V_{nm}^{ls,h} \left[j_\mu^p(\bm{k})\right]_{nm} \left[\delta\Delta_\nu^\dagger(\bm{k})\right]_{ls} \\ & + S_{nm}^{ls,p} \left[j_\mu^h(-\bm{k})\right]_{nm} \left[\delta\Delta_\nu(\bm{k})\right]_{ls} + S_{nm}^{ls,h} \left[j_\mu^h(-\bm{k})\right]_{nm} \left[\delta\Delta_\nu^\dagger(\bm{k})\right]_{ls} \Big), \end{aligned} \tag{S.33}$$

where $D_{\mu\nu}^{\text{geom1}}$ corresponds to the contribution from the quantum metric, $D_{\mu\nu}^{\text{geom2}}$ arises from interband pairing, $D_{\mu\nu}^{\text{multi}}$ represents the multi-gap contribution, and $D_{\mu\nu}^{\text{gap2}}$ captures the contribution from the momentum dependence of the gap function.

Next, we derive the explicit form of the coefficients $w_{in}^p$ and $w_{in}^h$. Let $\mathcal{G}(\bm{k})$ be the unitary matrix that diagonalizes the BdG Hamiltonian $\mathcal{H}_{\text{BdG}}(\bm{k})$. Then,

$$\mathcal{G}^\dagger(\bm{k}) \mathcal{H}_{\text{BdG}}(\bm{k}) \mathcal{G}(\bm{k}) = \begin{pmatrix} \bm{E}(\bm{k}) & 0 \\ 0 & -\bm{E}(\bm{k}) \end{pmatrix} = \begin{pmatrix} \text{diag}\big(E_1(\bm{k}), \cdots, E_f(\bm{k})\big) & 0 \\ 0 & -\text{diag}\big(E_1(\bm{k}), \cdots, E_f(\bm{k})\big) \end{pmatrix}, \tag{S.34}$$

where $E_i(\bm{k})$ are the positive quasiparticle energies in the superconducting state. Similarly, let $\mathcal{G}_0(\bm{k})$ be the unitary matrix that diagonalizes the normal-state Hamiltonian $\mathcal{H}_0(\bm{k})$. Then,

$$\mathcal{G}_0^\dagger(\bm{k}) \mathcal{H}_0(\bm{k}) \mathcal{G}_0(\bm{k}) = \bm{\varepsilon}(\bm{k}) = \text{diag}\big(\varepsilon_1(\bm{k}), \cdots, \varepsilon_f(\bm{k})\big), \tag{S.35}$$

where $\varepsilon_n(\bm{k})$ are the eigenenergies in the normal state. Using these unitary matrices, we can change the basis of the BdG Hamiltonian $\mathcal{H}_{\text{BdG}}(\bm{k})$. The transformed BdG Hamiltonian in the band basis is given by

$$\mathcal{H}_{\text{band}}(\bm{k}) := \begin{pmatrix} \mathcal{G}_0^\dagger(\bm{k}) & 0 \\ 0 & \mathcal{G}_0^\dagger(\bm{k}) \end{pmatrix} \mathcal{H}_{\text{BdG}}(\bm{k}) \begin{pmatrix} \mathcal{G}_0(\bm{k}) & 0 \\ 0 & \mathcal{G}_0(\bm{k}) \end{pmatrix} = \begin{pmatrix} \bm{\varepsilon}(\bm{k}) & \Delta(\bm{k}) \bm{1}_{f \times f} \\ \Delta^*(\bm{k}) \bm{1}_{f \times f} & -\bm{\varepsilon}(\bm{k}) \end{pmatrix}. \tag{S.36}$$

From the definition of the coefficients $w_{in}^p$ and $w_{in}^h$, we can express

$$\begin{pmatrix} w^p & w^h \end{pmatrix} = \mathcal{G}^\dagger(\bm{k}) \begin{pmatrix} \mathcal{G}_0(\bm{k}) & 0 \\ 0 & \mathcal{G}_0(\bm{k}) \end{pmatrix}. \tag{S.37}$$

Using this relation, the diagonal form of the BdG Hamiltonian can also be written as

$$\begin{pmatrix} \bm{E}(\bm{k}) & 0 \\ 0 & -\bm{E}(\bm{k}) \end{pmatrix} = \mathcal{G}^\dagger(\bm{k}) \mathcal{H}_{\text{BdG}}(\bm{k}) \mathcal{G}(\bm{k}) = \begin{pmatrix} w^p & w^h \end{pmatrix} \mathcal{H}_{\text{band}}(\bm{k}) \begin{pmatrix} w^{p\dagger} \\ w^{h\dagger} \end{pmatrix}. \tag{S.38}$$

Therefore, to determine $w_{in}^p$ and $w_{in}^h$, it is sufficient to diagonalize the BdG Hamiltonian of the band bases $\mathcal{H}_{\text{band}}(\bm{k})$.

To proceed, we introduce the unit vectors $e_i$, defined by $(e_i)_j = \delta_{ij}$, and construct the unitary matrix

$$U := (e_1, \, e_{f+1}, \, e_2, \, e_{f+2}, \, \cdots, \, e_f, \, e_{2f}). \tag{S.39}$$

Using this matrix, the band-basis BdG Hamiltonian is transformed into a block-diagonal form:

$$U^\dagger \mathcal{H}_{\text{band}}(\bm{k}) U = \begin{pmatrix} \varepsilon_1(\bm{k}) & \Delta(\bm{k}) & 0 & \cdots & 0 & 0 \\ \Delta^*(\bm{k}) & -\varepsilon_1(\bm{k}) & 0 & \cdots & 0 & 0 \\ 0 & 0 & \varepsilon_2(\bm{k}) & \cdots & 0 & 0 \\ \vdots & \vdots & \vdots & \ddots & \vdots & \vdots \\ 0 & 0 & 0 & \cdots & \varepsilon_f(\bm{k}) & \Delta(\bm{k}) \\ 0 & 0 & 0 & \cdots & \Delta^*(\bm{k}) & -\varepsilon_f(\bm{k}) \end{pmatrix}. \tag{S.40}$$

Since $U^\dagger \mathcal{H}_{\text{band}}(\bm{k}) U$ is block-diagonal with $f$ $2 \times 2$ blocks, we can diagonalize each block separately. Let $\mathcal{U}_i(\bm{k})$ denote the unitary matrix that diagonalizes the $i$-th $2 \times 2$ block. Then,

$$\mathcal{U}_i^\dagger(\bm{k}) \begin{pmatrix} \varepsilon_i(\bm{k}) & \Delta(\bm{k}) \\ \Delta^*(\bm{k}) & -\varepsilon_i(\bm{k}) \end{pmatrix} \mathcal{U}_i(\bm{k}) = \begin{pmatrix} E_i(\bm{k}) & 0 \\ 0 & -E_i(\bm{k}) \end{pmatrix}, \tag{S.41}$$

$$\mathcal{U}_i(\bm{k}) = \begin{pmatrix} u_i(\bm{k}) & -v_i(\bm{k}) \\ v_i^*(\bm{k}) & u_i(\bm{k}) \end{pmatrix}, \quad E_i(\bm{k}) = \sqrt{\varepsilon_i(\bm{k})^2 + |\Delta(\bm{k})|^2}, \tag{S.42}$$

$$u_i(\bm{k}) = \frac{1}{\sqrt{2}} \sqrt{1 + \frac{\varepsilon_i(\bm{k})}{E_i(\bm{k})}}, \quad v_i(\bm{k}) = \frac{1}{\sqrt{2}} \frac{\Delta(\bm{k})}{\sqrt{E_i(\bm{k})(E_i(\bm{k}) + \varepsilon_i(\bm{k}))}}. \tag{S.43}$$

Using the diagonalizing matrices $\mathcal{U}_i(\bm{k})$, we define the block-diagonal unitary matrix $\mathcal{U}(\bm{k})$ as

$$\mathcal{U}(\bm{k}) = \begin{pmatrix} \mathcal{U}_1(\bm{k}) & 0 & \cdots & 0 \\ 0 & \mathcal{U}_2(\bm{k}) & \cdots & 0 \\ \vdots & \vdots & \ddots & \vdots \\ 0 & 0 & \cdots & \mathcal{U}_f(\bm{k}) \end{pmatrix}. \tag{S.44}$$

Next, we define the permutation matrix $V$ as

$$V = (e_1, \, e_3, \, \cdots, \, e_{2f-1}, \, e_2, \, e_4, \, \cdots, \, e_{2f}). \tag{S.45}$$

This permutation rearranges the order of the eigenstates to group all positive-energy and negative-energy components together. Combining all transformations, the full diagonalization of the band-basis BdG Hamiltonian becomes

$$V^\dagger \mathcal{U}^\dagger(\bm{k}) U^\dagger \mathcal{H}_{\text{band}}(\bm{k}) U \mathcal{U}(\bm{k}) V = V^\dagger \begin{pmatrix} E_1(\bm{k}) & 0 & \cdots & 0 & 0 \\ 0 & -E_1(\bm{k}) & \cdots & 0 & 0 \\ \vdots & \vdots & \ddots & \vdots & \vdots \\ 0 & 0 & \cdots & E_f(\bm{k}) & 0 \\ 0 & 0 & \cdots & 0 & -E_f(\bm{k}) \end{pmatrix} V = \begin{pmatrix} \bm{E}(\bm{k}) & 0 \\ 0 & -\bm{E}(\bm{k}) \end{pmatrix}. \tag{S.46}$$

By evaluating this transformation, we obtain

$$\begin{pmatrix} w^p & w^h \end{pmatrix} = V^\dagger \mathcal{U}^\dagger(\bm{k}) U^\dagger$$
$$= \begin{pmatrix} u_1(\bm{k}) & 0 & \cdots & 0 & v_1(\bm{k}) & 0 & \cdots & 0 \\ 0 & u_2(\bm{k}) & \cdots & 0 & 0 & v_2(\bm{k}) & \cdots & 0 \\ \vdots & \vdots & \ddots & \vdots & \vdots & \vdots & \ddots & \vdots \\ 0 & 0 & \cdots & u_f(\bm{k}) & 0 & 0 & \cdots & v_f(\bm{k}) \\ -v_1^*(\bm{k}) & 0 & \cdots & 0 & u_1(\bm{k}) & 0 & \cdots & 0 \\ 0 & -v_2^*(\bm{k}) & \cdots & 0 & 0 & u_2(\bm{k}) & \cdots & 0 \\ \vdots & \vdots & \ddots & \vdots & \vdots & \vdots & \ddots & \vdots \\ 0 & 0 & \cdots & -v_f^*(\bm{k}) & 0 & 0 & \cdots & u_f(\bm{k}) \end{pmatrix}. \tag{S.47}$$

Using the explicit expressions for $w_{in}^p$ and $w_{in}^h$, we can now evaluate each term in the decomposition of the superfluid weight. The conventional and quantum-geometric contributions are given by

$$D_{\mu\nu}^{\text{conv1}} = \int_{\text{BZ}} \frac{d^d \boldsymbol{k}}{(2\pi)^d} \sum_n \left( 2f'(E_n(\boldsymbol{k})) + \frac{f(-E_n(\boldsymbol{k})) - f(E_n(\boldsymbol{k}))}{E_n(\boldsymbol{k})} \right) \frac{|\Delta(\boldsymbol{k})|^2}{E_n^2(\boldsymbol{k})} \partial_\mu \varepsilon_n(\boldsymbol{k}) \partial_\nu \varepsilon_n(\boldsymbol{k}), \tag{S.48}$$

$$D_{\mu\nu}^{\text{gap1}} = -\int_{\text{BZ}} \frac{d^d \boldsymbol{k}}{(2\pi)^d} \sum_n \left( 2f'(E_n(\boldsymbol{k})) + \frac{f(-E_n(\boldsymbol{k})) - f(E_n(\boldsymbol{k}))}{E_n(\boldsymbol{k})} \right) \text{Re}\left[ \frac{\Delta^*(\boldsymbol{k}) \varepsilon_n(\boldsymbol{k})}{E_n^2(\boldsymbol{k})} \partial_\mu \varepsilon_n(\boldsymbol{k}) \partial_\nu \Delta(\boldsymbol{k}) \right], \tag{S.49}$$

$$D_{\mu\nu}^{\text{geom1}} = \int_{\text{BZ}} \frac{d^d \boldsymbol{k}}{(2\pi)^d} \sum_{n\neq m} \left( \frac{f(-E_m) - f(E_m)}{E_m} - \frac{f(-E_n) - f(E_n)}{E_n} \right) \frac{|\Delta(\boldsymbol{k})|^2}{E_n^2(\boldsymbol{k}) - E_m^2(\boldsymbol{k})} (\varepsilon_n(\boldsymbol{k}) - \varepsilon_m(\boldsymbol{k}))^2 g_{\mu\nu}^{nm}(\boldsymbol{k}). \tag{S.50}$$

In the quantum-geometric contribution $D_{\mu\nu}^{\text{geom1}}$, we see the band-resolved quantum metric defined by

$$g_{\mu\nu}^{nm}(\boldsymbol{k}) \coloneqq 2\, \text{Re}\, \langle u_n | \partial_\mu u_m \rangle \langle \partial_\nu u_m | u_n \rangle. \tag{S.51}$$

The remaining contributions, $D_{\mu\nu}^{\text{geom2}}$, $D_{\mu\nu}^{\text{multi}}$, and $D_{\mu\nu}^{\text{gap2}}$, vanish under the assumption of single-gap intra-band pairing. Thus, the total conventional and quantum-geometric contributions are reduced to

$$\begin{aligned}
D_{\mu\nu}^{\text{conv}}(T) = \int_{\text{BZ}} \frac{d^d \boldsymbol{k}}{(2\pi)^d} \sum_n & \left( 2f'(E_n(\boldsymbol{k})) + \frac{f(-E_n(\boldsymbol{k})) - f(E_n(\boldsymbol{k}))}{E_n(\boldsymbol{k})} \right) \\
& \times \text{Re}\left[ \frac{\Delta^*(\boldsymbol{k}) \partial_\mu \varepsilon_n(\boldsymbol{k})}{E_n^2(\boldsymbol{k})} \left( \Delta(\boldsymbol{k}) \partial_\nu \varepsilon_n(\boldsymbol{k}) - \varepsilon_n(\boldsymbol{k}) \partial_\nu \Delta(\boldsymbol{k}) \right) \right],
\end{aligned} \tag{S.52}$$

$$\begin{aligned}
D_{\mu\nu}^{\text{geom}}(T) = \int_{\text{BZ}} \frac{d^d \boldsymbol{k}}{(2\pi)^d} \sum_{n\neq m} & \left( \frac{f(-E_m) - f(E_m)}{E_m} - \frac{f(-E_n) - f(E_n)}{E_n} \right) \\
& \times \frac{|\Delta(\boldsymbol{k})|^2}{E_n^2(\boldsymbol{k}) - E_m^2(\boldsymbol{k})} (\varepsilon_n(\boldsymbol{k}) - \varepsilon_m(\boldsymbol{k}))^2 g_{\mu\nu}^{nm}(\boldsymbol{k}).
\end{aligned} \tag{S.53}$$

## TEMPERATURE DEPENDENCE OF THE QUANTUM-GEOMETRIC SUPERFLUID WEIGHT

We describe the procedure for calculating the temperature dependence of the superfluid weight arising from quantum geometry. We begin by restating the formula of the quantum-geometric contribution to the superfluid weight Eq. (S.53):

$$\begin{aligned}
D_{\mu\nu}^{\text{geom}}(T) = \int_{\text{BZ}} \frac{d^d \boldsymbol{k}}{(2\pi)^d} \sum_{n\neq m} & \left( \frac{f(-E_m) - f(E_m)}{E_m} - \frac{f(-E_n) - f(E_n)}{E_n} \right) \\
& \times \frac{|\Delta(\boldsymbol{k})|^2}{E_n^2(\boldsymbol{k}) - E_m^2(\boldsymbol{k})} (\varepsilon_n(\boldsymbol{k}) - \varepsilon_m(\boldsymbol{k}))^2 g_{\mu\nu}^{nm}(\boldsymbol{k}).
\end{aligned} \tag{S.53}$$

Since we focus on the low-temperature behavior, the temperature dependence of the gap function is neglected. We then study the following quantity:

$$\delta D_{\mu\nu}^{\text{geom}}(T) \coloneqq D_{\mu\nu}^{\text{geom}}(T=0) - D_{\mu\nu}^{\text{geom}}(T), \tag{S.54}$$

which represents the reduction in the quantum-geometric superfluid weight at temperature $T$ relative to its zero-temperature value. Differentiating with respect to temperature yields

$$\frac{\partial \delta D_{\mu\nu}^{\text{geom}}(T)}{\partial T} = \sum_{n\neq m} \frac{2}{T} \int_{\text{BZ}} \frac{d^d \boldsymbol{k}}{(2\pi)^d} \left( f'(E_m) - f'(E_n) \right) \frac{|\Delta(\boldsymbol{k})|^2}{E_m^2(\boldsymbol{k}) - E_n^2(\boldsymbol{k})} (\varepsilon_n(\boldsymbol{k}) - \varepsilon_m(\boldsymbol{k}))^2 g_{\mu\nu}^{nm}(\boldsymbol{k}). \tag{S.55}$$

This can be rewritten in terms of the density of states as

$$\frac{\partial \delta D_{\mu\nu}^{\text{geom}}(T)}{\partial T} = -\sum_{n\neq m} \frac{2}{T} \int_0^\infty dE\, f'(E) \left[ D_n(E) \left\langle \frac{|\Delta(\boldsymbol{k})|^2}{E_m^2(\boldsymbol{k}) - E_n^2(\boldsymbol{k})} (\varepsilon_n(\boldsymbol{k}) - \varepsilon_m(\boldsymbol{k}))^2 g_{\mu\nu}^{nm}(\boldsymbol{k}) \right\rangle_{E,n} + (n \leftrightarrow m) \right]. \tag{S.56}$$

In the case where band $n$ is flat and the gap function is momentum independent, the expression reduces to

$$\frac{\partial \delta D_{\mu\nu}^{\text{geom}}(T)}{\partial T} = \sum_{m\neq n} \frac{2}{T} \int_0^\infty \mathrm{d}E \Big(f'(E) - f'(|\Delta|)\Big) D_m(E) \left\langle \frac{|\Delta|^2}{\varepsilon_m^2(\boldsymbol{k})} \big(\varepsilon_n(\boldsymbol{k}) - \varepsilon_m(\boldsymbol{k})\big)^2 g_{\mu\nu}^{nm}(\boldsymbol{k}) \right\rangle_{E,m} + \cdots, \quad (\text{S.57})$$

where the ellipsis stands for contributions from bands not involving $n$, which are omitted here for simplicity. In addition, if both bands $n$ and $m$ lie at the Fermi level and satisfy $E_n(\boldsymbol{k}) = E_m(\boldsymbol{k})$, we find

$$\frac{\partial \delta D_{\mu\nu}^{\text{geom}}(T)}{\partial T} = \frac{1}{T} \int_0^\infty \mathrm{d}E f''(E) D_n(E) \left\langle \frac{|\Delta(\boldsymbol{k})|^2}{E_n(\boldsymbol{k})} \big(\varepsilon_n(\boldsymbol{k}) - \varepsilon_m(\boldsymbol{k})\big)^2 g_{\mu\nu}^{nm}(\boldsymbol{k}) \right\rangle_{E,n} + \cdots, \quad (\text{S.58})$$

where the omitted terms originate from bands other than $n$ and $m$. In the above expressions, $D_n(E)$ and $\langle \mathcal{O}(\boldsymbol{k}) \rangle_{E,n}$ respectively denote the density of states and the expectation value of $\mathcal{O}(\boldsymbol{k})$ for band $n$ at energy $E$, defined as

$$D_n(E) := \int_{\text{BZ}} \frac{\mathrm{d}^d \boldsymbol{k}}{(2\pi)^d} \delta\big(E - E_n(\boldsymbol{k})\big), \quad (\text{S.59})$$

$$\langle \mathcal{O}(\boldsymbol{k}) \rangle_{E,n} := \frac{1}{D_n(E)} \int_{\text{BZ}} \frac{\mathrm{d}^d \boldsymbol{k}}{(2\pi)^d} \mathcal{O}(\boldsymbol{k}) \delta\big(E - E_n(\boldsymbol{k})\big). \quad (\text{S.60})$$

To determine the leading-order behavior of the quantum-geometric superfluid weight at low temperatures, it is sufficient to evaluate the low-energy asymptotics of Eqs. (S.59) and (S.60). In what follows, we first compute the density of states, followed by the evaluation of the energy-resolved expectation values.

### I. Density of States

The density of states $D_n(E)$ quantifies the number of states for band $n$ at energy $E$. Here, we compute the density of states for a two-dimensional flat band at positive energies ($E \geq 0$).

#### A. Point Nodes

We first consider the case of point nodes. Near a point node, the energy dispersion of Bogoliubov quasiparticles is approximated as

$$E_n(\boldsymbol{p}) = \sqrt{\alpha_1^2 \Delta^2 p_1^{2l} + \alpha_2^2 \Delta^2 p_2^{2s}}, \quad (\text{S.61})$$

where $\boldsymbol{p} = (p_1, p_2)$ denotes the momentum relative to the nodal point, and $\alpha_1, \alpha_2$ are constants. The integers $l$ and $s$ characterize the power-law dispersion along the $p_1$ and $p_2$ directions, respectively [3]. The density of states is then given by

$$\begin{aligned} D_n(E) &= \int_{\text{BZ}} \frac{\mathrm{d}^2 \boldsymbol{k}}{(2\pi)^2} \delta(E - E_n(\boldsymbol{k})) \simeq \int_{-\infty}^{\infty} \frac{\mathrm{d}^2 \boldsymbol{p}}{(2\pi)^2} \delta\bigg(E - \sqrt{\alpha_1^2 \Delta^2 p_1^{2l} + \alpha_2^2 \Delta^2 p_2^{2s}}\bigg) \\ &= \frac{4}{(2\pi)^2} \frac{1}{ls(\alpha_1 \Delta)^{1/l}(\alpha_2 \Delta)^{1/s}} \int_0^\infty \mathrm{d}^2 \boldsymbol{p} \, p_1^{1/l-1} p_2^{1/s-1} \delta\bigg(E - \sqrt{p_1^2 + p_2^2}\bigg) \\ &= \frac{2}{(2\pi)^2} \frac{1}{ls(\alpha_1 \Delta)^{1/l}(\alpha_2 \Delta)^{1/s}} \frac{\Gamma(1/2l)\Gamma(1/2s)}{\Gamma(1/2l + 1/2s)} E^{1/l+1/s-1} \propto E^{1/l+1/s-1}. \end{aligned} \quad (\text{S.62})$$

Here, $\Gamma(x)$ denotes the Gamma function.

#### B. Line Nodes without Crossing

We next consider the case of a line node that does not intersect any other line nodes on the Fermi surface. In the vicinity of such a line node, the energy dispersion of Bogoliubov quasiparticles is approximated as

$$E_n(\boldsymbol{p}) = \sqrt{\alpha_1^2 \Delta^2 p_1^{2l}}, \quad (\text{S.63})$$

where $p_1$ denotes the momentum perpendicular to the line node. The corresponding density of states can be evaluated as

$$D_n(E) = \int_{\text{BZ}} \frac{d^2\bm{k}}{(2\pi)^2} \delta(E - E_n(\bm{k})) \simeq \int_0^L \frac{dp_2}{2\pi} \int_{-\infty}^{\infty} \frac{dp_1}{2\pi} \delta\left(E - \sqrt{\alpha_1^2 \Delta^2 p_1^{2l}}\right)$$
$$= \frac{L}{2\pi} \frac{1}{\pi l(\alpha_1 \Delta)^{1/l}} \int_0^{\infty} dp_1 p_1^{1/l - 1} \delta(E - p_1) = \frac{L}{2\pi^2} \frac{1}{l(\alpha_1 \Delta)^{1/l}} E^{1/l - 1} \propto E^{1/l - 1}, \quad \text{(S.64)}$$

where $L$ denotes the length of the line node.

### C. Line Nodes with Crossing

We now analyze the density of states associated with intersecting line nodes on the Fermi surface. Near the intersection point, the energy dispersion of Bogoliubov quasiparticles can be approximated as

$$E_n(\bm{p}) = \sqrt{\alpha_1^2 \Delta^2 p_1^{2l} p_2^{2l}}, \quad \text{(S.65)}$$

where $\bm{p} = (p_1, p_2)$ denotes the momentum measured relative to the nodal intersection point. The density of states in this case is given by

$$D_n(E) = \int_{\text{BZ}} \frac{d^2\bm{k}}{(2\pi)^2} \delta(E - E_n(\bm{k})) \simeq \int_0^L \frac{d^2\bm{p}}{(2\pi)^2} \delta\left(E - \sqrt{\alpha_1^2 \Delta^2 p_1^{2l} p_2^{2l}}\right)$$
$$= \frac{1}{(2\pi)^2} \frac{1}{l^2(\alpha_1 \Delta)^{1/l}} \int_0^{\alpha_1 \Delta L^{2l}} dp_1 \int_{p_1/\alpha_1 \Delta L^l}^{L^l} dp_2 p_1^{1/l - 1} p_2^{-1} \delta(E - p_1) \quad \text{(S.66)}$$
$$= \frac{1}{(2\pi)^2} \frac{1}{l^2(\alpha_1 \Delta)^{1/l}} E^{1/l - 1} \ln\left(\frac{\alpha_1 \Delta L^{2l}}{E}\right) \propto -E^{1/l - 1} \ln E.$$

In the last expression, we have extracted the low-energy behavior in the leading order.

### D. General Results

| gap type | dispersive band | flat band | Dirac band |
|---|---|---|---|
| full gap | $E(E^2 - \Delta^2)^{-1/2} \Theta(E - \Delta)$ | $\delta(E - \Delta)$ | $E\Theta(E - \Delta)$ |
| point nodes | $E^{1/l + 1/s}$ | $E^{1/l + 1/s - 1}$ | $E$ |
| line nodes (no crossing) | $E^{1/l}$ | $E^{1/l - 1}$ | $E$ |
| line nodes (with crossing) | $-E^{1/l} \ln E$ | $-E^{1/l - 1} \ln E$ | $E$ |

Supplementary Table S.I. Energy dependence of the density of states for various band structures and gap structures. For point nodes and non-crossing line nodes in dispersive and flat bands, $l$ and $s$ denote arbitrary positive integers characterizing the order of dispersion of the nodal gap. For crossing line nodes, $l, s = 1$ or $2$. In the case of Dirac bands, nodal gaps with linear or quadratic dispersion are considered. $\Theta(x)$ denotes the Heaviside step function.

By performing similar calculations for the remaining cases, we obtain the results summarized in Table S.I. The dispersive band refers to a conventional band structure where the normal-state energy dispersion near the Fermi surface can be expressed as

$$\varepsilon_n(\bm{p}) = v_\text{F} p_3. \quad \text{(S.67)}$$

Here, $p_3$ represents the momentum normal to the Fermi surface and $v_\text{F}$ denotes the Fermi velocity. The flat band, for which some gap structures are discussed above, is assumed to be two-dimensional with $v_\text{F} = 0$. The Dirac band is characterized by a normal-state dispersion of the form

$$\varepsilon_n(\bm{p}) = v_\text{F} \sqrt{p_1^2 + p_2^2}. \quad \text{(S.68)}$$

## II. Expectation Values

We evaluate the expectation value that appear in Eq. (S.56) for two cases: in the absence and presence of band crossing at the Fermi level. Throughout the following analysis, we assume that the quantity $\bigl(\varepsilon_n(\bm{k}) - \varepsilon_m(\bm{k})\bigr)^2 g^{nm}_{\mu\nu}(\bm{k})$ is finite at the nodal points.

### A. Cases in the Absence of Band Crossing at the Fermi Level

First, we consider the case where there is no band crossing at the Fermi level. In this case, it is sufficient to study the system with only the band $n$ crossing the Fermi level. The contribution of other band $m$ can be estimated by considering the quantity

$$W_m := \min_{\bm{k}_\mathrm{F}} \Bigl( \sqrt{\varepsilon_m^2(\bm{k}) + |\Delta(\bm{k})|^2} - |\Delta(\bm{k})| \Bigr), \tag{S.69}$$

which represents the minimum excitation energy of band $m$ relative to the superconducting gap. Since the derivative of the Fermi distribution function $f'(E_m)$ is exponentially suppressed compared to $f'(E_n)$ as $\sim e^{-W_m/T}$, we can neglect the contribution of the band $m$ in the low-temperature regime $T \ll W_m$. Therefore, the expectation value can be evaluated by considering only the band $n$. Let us define $W_{nm}(\bm{k}) := \sqrt{\varepsilon_m^2(\bm{k}) - \varepsilon_n^2(\bm{k})}$, which remains strictly positive around the Fermi surface. To exemplify the general procedure, we assume the case where the flat band $n$ has a point node at momentum $\bm{k}_\mathrm{n}$. In the low-energy limit, the leading-order contribution to the expectation value becomes

$$\begin{aligned}
&\left\langle \frac{|\Delta(\bm{k})|^2}{E_m^2(\bm{k}) - E_n^2(\bm{k})} \bigl(\varepsilon_n(\bm{k}) - \varepsilon_m(\bm{k})\bigr)^2 g^{nm}_{\mu\nu}(\bm{k}) \right\rangle_{E,n} \\
&= \frac{1}{D_n(E)} \int_{\mathrm{BZ}} \frac{\mathrm{d}^2\bm{k}}{(2\pi)^2} \delta\bigl(E - E_n(\bm{k})\bigr) \frac{|\Delta(\bm{k})|^2}{W_{nm}^2(\bm{k})} \bigl(\varepsilon_n(\bm{k}) - \varepsilon_m(\bm{k})\bigr)^2 g^{nm}_{\mu\nu}(\bm{k}) \\
&\simeq \frac{1}{D_n(E)} \int_{-\infty}^{\infty} \frac{\mathrm{d}^2\bm{p}}{(2\pi)^2} \delta\bigl(E - E_n(\bm{p})\bigr) \frac{|\Delta(\bm{p})|^2}{W_{nm}^2(\bm{k}_\mathrm{n}) + \mathcal{O}(|\bm{p}|)} \Bigl( \varepsilon_m^2(\bm{k}_\mathrm{n}) g^{nm}_{\mu\nu}(\bm{k}_\mathrm{n}) + \mathcal{O}(|\bm{p}|) \Bigr) \\
&= \frac{1}{D_n(E)} \frac{4}{(2\pi)^2} \frac{1}{ls(\alpha_1 \Delta)^{1/l}(\alpha_2 \Delta)^{1/s}} \\
&\quad \times \int_0^\infty \mathrm{d}p \int_0^{\pi/2} \mathrm{d}\theta\, p^{1/l+1/s-1} \cos^{1/l-1}\theta \sin^{1/s-1}\theta\, \delta(E-p) p^2 \left( \frac{1}{W_{nm}^2(\bm{k}_\mathrm{n})} \varepsilon_m^2(\bm{k}_\mathrm{n}) g^{nm}_{\mu\nu}(\bm{k}_\mathrm{n}) + \mathcal{O}(E^{1/t}) \right) \\
&= \left\langle \frac{\varepsilon_m^2(\bm{k}) g^{nm}_{\mu\nu}(\bm{k})}{W_{nm}^2(\bm{k})} \right\rangle_{E=0,n} E^2 + \mathcal{O}(E^{2+1/t}) \propto E^2,
\end{aligned} \tag{S.70}$$

where $t = \min(l, s)$ and the final expression shows the leading-order behavior in the low-energy region. The expectation value at zero energy is evaluated as

$$\begin{aligned}
\left\langle \frac{\varepsilon_m^2(\bm{k}) g^{nm}_{\mu\nu}(\bm{k})}{W_{nm}^2(\bm{k})} \right\rangle_{E=0,n}
&= \frac{1}{D_n(E=0)} \int_{\mathrm{BZ}} \frac{\mathrm{d}^2\bm{k}}{(2\pi)^2} \delta\bigl(E_n(\bm{k})\bigr) \frac{\varepsilon_m^2(\bm{k}) g^{nm}_{\mu\nu}(\bm{k})}{W_{nm}^2(\bm{k})} \\
&= \frac{1}{D_n(E=0)} \int_{\mathrm{BZ}} \frac{\mathrm{d}^2\bm{k}}{(2\pi)^2} \delta\bigl(E_n(\bm{k})\bigr) \frac{\varepsilon_m^2(\bm{k}_\mathrm{n}) g^{nm}_{\mu\nu}(\bm{k}_\mathrm{n})}{W_{nm}^2(\bm{k}_\mathrm{n})} \\
&= \frac{1}{D_n(E=0)} D_n(E=0) \frac{\varepsilon_m^2(\bm{k}_\mathrm{n}) g^{nm}_{\mu\nu}(\bm{k}_\mathrm{n})}{W_{nm}^2(\bm{k}_\mathrm{n})} \\
&= \frac{\varepsilon_m^2(\bm{k}_\mathrm{n}) g^{nm}_{\mu\nu}(\bm{k}_\mathrm{n})}{W_{nm}^2(\bm{k}_\mathrm{n})}.
\end{aligned} \tag{S.71}$$

The same analysis can be applied to other gap structures and band structures. The low-energy behaviors of the expectation value in the leading order are summarized in Table S.II.

### B. Cases in the Presence of Band Crossing at the Fermi Level

Next, we consider the case where bands $n$ and $m$ cross at the Fermi level. We assume that the quantity $\bigl(\varepsilon_n(\bm{k}) - \varepsilon_m(\bm{k})\bigr)^2 g^{nm}_{\mu\nu}(\bm{k})$ remains finite at the band degeneracy point. We analyze the following representative models with band de-

9| gap type | dispersive band | flat band |
|---|---|---|
| full gap | const. | const. |
| point nodes | $E^2$ | $E^2$ |
| line nodes (no crossing) | $E^2$ | $E^2$ |
| line nodes (with crossing) | $E^2$ | $E^2$ |

Supplementary Table S.II. Leading-order low-energy behavior of the energy-resolved expectation value that appear in Eq. (S.56) for various gap structures. We assume dispersive or flat band structures without band crossing at the Fermi level. The results apply to arbitrary order of dispersion of the nodal gap.

generacy:

(a) A flat band and a Dirac band

$$\varepsilon_n(\boldsymbol{p}) = 0, \tag{S.72}$$
$$\varepsilon_m(\boldsymbol{p}) = v_\mathrm{F}\sqrt{p_1^2 + p_2^2}, \tag{S.73}$$

where $\boldsymbol{p}$ denotes the momentum relative to the Dirac point.

(b) The Dirac band

$$\varepsilon_n(\boldsymbol{p}) = v_\mathrm{F}\sqrt{p_1^2 + p_2^2}, \tag{S.74}$$
$$\varepsilon_m(\boldsymbol{p}) = -v_\mathrm{F}\sqrt{p_1^2 + p_2^2}. \tag{S.75}$$

(c) Two dispersive bands

$$\varepsilon_n(\boldsymbol{p}) = v_\mathrm{F} p_3, \tag{S.76}$$
$$\varepsilon_m(\boldsymbol{p}) = -v_\mathrm{F} p_3, \tag{S.77}$$

where $p_3$ is the momentum normal to the Fermi surface.

In the models (b) and (c), $E_n(\boldsymbol{k}) = E_m(\boldsymbol{k})$, and we need to evaluate the expectation value,

$$\left\langle \frac{|\Delta(\boldsymbol{k})|^2}{E_n(\boldsymbol{k})} \big(\varepsilon_n(\boldsymbol{k}) - \varepsilon_m(\boldsymbol{k})\big)^2 g_{\mu\nu}^{nm}(\boldsymbol{k}) \right\rangle_{E,n}, \tag{S.78}$$

which appear in Eq. (11) [see the main text]. To show examples of explicit calculations, we evaluate the leading-order asymptotic behaviors of the expectation values for the two cases: (1) A linear point node lies at the degenerate momentum between a flat band and a Dirac band in the model (a). (2) A point node lies at the Dirac point in the model (b).

First, we compute the expectation value associated with the band $n$, i.e. the flat band, in case (1). This yields

$$\begin{aligned}
&\left\langle \frac{|\Delta(\boldsymbol{k})|^2}{E_m^2(\boldsymbol{k}) - E_n^2(\boldsymbol{k})} \big(\varepsilon_n(\boldsymbol{k}) - \varepsilon_m(\boldsymbol{k})\big)^2 g_{\mu\nu}^{nm}(\boldsymbol{k}) \right\rangle_{E,n} \\
&= \frac{1}{D_n(E)} \int_\mathrm{BZ} \frac{\mathrm{d}^2 \boldsymbol{k}}{(2\pi)^2} \delta\big(E - E_n(\boldsymbol{k})\big) \frac{|\Delta(\boldsymbol{k})|^2}{\varepsilon_m^2(\boldsymbol{k})} \varepsilon_m^2(\boldsymbol{k}) g_{\mu\nu}^{nm}(\boldsymbol{k}) \\
&\simeq \frac{1}{D_n(E)} \int_{-\infty}^{\infty} \frac{\mathrm{d}^2 \boldsymbol{p}}{(2\pi)^2} \delta\big(E - E_n(\boldsymbol{p})\big) \frac{|\Delta(\boldsymbol{p})|^2}{\varepsilon_m^2(\boldsymbol{p})} \Big(\varepsilon_m^2(\boldsymbol{k}_\mathrm{n}) g_{\mu\nu}^{nm}(\boldsymbol{k}_\mathrm{n}) + \mathcal{O}(|\boldsymbol{p}|)\Big) \\
&= \frac{1}{D_n(E)} \frac{4}{(2\pi)^2} \frac{1}{\alpha_1 \alpha_2 \Delta^2} \int_0^\infty \mathrm{d}^2 \boldsymbol{p}\, \delta\!\left(E - \sqrt{p_1^2 + p_2^2}\right) \\
&\quad \times \frac{p_1^2 + p_2^2}{v_\mathrm{F}^2} \left(\frac{p_1^2}{\alpha_1^2 \Delta^2} + \frac{p_2^2}{\alpha_2^2 \Delta^2}\right)^{-1} \Big(\varepsilon_m^2(\boldsymbol{k}_\mathrm{n}) g_{\mu\nu}^{nm}(\boldsymbol{k}_\mathrm{n}) + \mathcal{O}(E)\Big) \\
&= \frac{\alpha_1 \alpha_2 \Delta^2}{v_\mathrm{F}^2} \left\langle \varepsilon_m^2(\boldsymbol{k}) g_{\mu\nu}^{nm}(\boldsymbol{k}) \right\rangle_{E=0, n} + \mathcal{O}(E).
\end{aligned} \tag{S.79}$$

Second, we compute the expectation value associated with the band $m$, i.e., the Dirac band:

$$\begin{aligned}
&\left\langle \frac{|\Delta(\boldsymbol{k})|^2}{E_n^2(\boldsymbol{k}) - E_m^2(\boldsymbol{k})} (\varepsilon_n(\boldsymbol{k}) - \varepsilon_m(\boldsymbol{k}))^2 g_{\mu\nu}^{nm}(\boldsymbol{k}) \right\rangle_{E,m} \\
&= -\frac{1}{D_m(E)} \int_{\mathrm{BZ}} \frac{\mathrm{d}^2\boldsymbol{k}}{(2\pi)^2} \delta(E - E_m(\boldsymbol{k})) \frac{|\Delta(\boldsymbol{k})|^2}{\varepsilon_m^2(\boldsymbol{k})} \varepsilon_m^2(\boldsymbol{k}) g_{\mu\nu}^{nm}(\boldsymbol{k}) \\
&\simeq -\frac{1}{D_m(E)} \int_{-\infty}^{\infty} \frac{\mathrm{d}^2\boldsymbol{p}}{(2\pi)^2} \delta(E - E_m(\boldsymbol{p})) \frac{|\Delta(\boldsymbol{p})|^2}{\varepsilon_m^2(\boldsymbol{p})} \left(\varepsilon_m^2(\boldsymbol{k}_{\mathrm{n}}) g_{\mu\nu}^{nm}(\boldsymbol{k}_{\mathrm{n}}) + \mathcal{O}(|\boldsymbol{p}|)\right) \\
&= -\frac{1}{D_m(E)} \frac{4}{(2\pi)^2} \frac{1}{\sqrt{(\alpha_1^2\Delta^2 + v_{\mathrm{F}}^2)(\alpha_2^2\Delta^2 + v_{\mathrm{F}}^2)}} \int_0^{\infty} \mathrm{d}^2\boldsymbol{p}\, \delta\left(E - \sqrt{p_1^2 + p_2^2}\right) \\
&\quad \times \frac{1}{v_{\mathrm{F}}^2} \left(\frac{\alpha_1^2\Delta^2 p_1^2}{\alpha_1^2\Delta^2 + v_{\mathrm{F}}^2} + \frac{\alpha_2^2\Delta^2 p_2^2}{\alpha_2^2\Delta^2 + v_{\mathrm{F}}^2}\right) \left(\frac{p_1^2}{\alpha_1^2\Delta^2 + v_{\mathrm{F}}^2} + \frac{p_2^2}{\alpha_2^2\Delta^2 + v_{\mathrm{F}}^2}\right)^{-1} \left(\varepsilon_m^2(\boldsymbol{k}_{\mathrm{n}}) g_{\mu\nu}^{nm}(\boldsymbol{k}_{\mathrm{n}}) + \mathcal{O}(E)\right) \\
&= -\frac{\alpha_1^2\Delta^2 \sqrt{\alpha_2^2\Delta^2 + v_{\mathrm{F}}^2} + \alpha_2^2\Delta^2 \sqrt{\alpha_1^2\Delta^2 + v_{\mathrm{F}}^2}}{v_{\mathrm{F}}^2 \left(\sqrt{\alpha_1^2\Delta^2 + v_{\mathrm{F}}^2} + \sqrt{\alpha_2^2\Delta^2 + v_{\mathrm{F}}^2}\right)} \left\langle \varepsilon_m^2(\boldsymbol{k}) g_{\mu\nu}^{nm}(\boldsymbol{k}) \right\rangle_{E=0,m} + \mathcal{O}(E).
\end{aligned} \tag{S.80}$$

Next, we turn to case (2). The expectation value is evaluated as

$$\begin{aligned}
&\left\langle \frac{|\Delta(\boldsymbol{k})|^2}{E_n(\boldsymbol{k})} (\varepsilon_n(\boldsymbol{k}) - \varepsilon_m(\boldsymbol{k}))^2 g_{\mu\nu}^{nm}(\boldsymbol{k}) \right\rangle_{E,n} \\
&= \frac{1}{D_n(E)E} \int_{\mathrm{BZ}} \frac{\mathrm{d}^2\boldsymbol{k}}{(2\pi)^2} \delta(E - E_n(\boldsymbol{k})) |\Delta(\boldsymbol{k})|^2 (\varepsilon_n(\boldsymbol{k}) - \varepsilon_m(\boldsymbol{k}))^2 g_{\mu\nu}^{nm}(\boldsymbol{k}) \\
&\simeq \frac{1}{D_n(E)E} \int_{-\infty}^{\infty} \frac{\mathrm{d}^2\boldsymbol{p}}{(2\pi)^2} \delta(E - E_n(\boldsymbol{p})) |\Delta(\boldsymbol{p})|^2 \left((\varepsilon_n(\boldsymbol{k}_{\mathrm{n}}) - \varepsilon_m(\boldsymbol{k}_{\mathrm{n}}))^2 g_{\mu\nu}^{nm}(\boldsymbol{k}_{\mathrm{n}}) + \mathcal{O}(|\boldsymbol{p}|)\right) \\
&= \frac{1}{D_n(E)E} \frac{4}{(2\pi)^2} \frac{1}{\sqrt{(\alpha_1^2\Delta^2 + v_{\mathrm{F}}^2)(\alpha_2^2\Delta^2 + v_{\mathrm{F}}^2)}} \int_0^{\infty} \mathrm{d}^2\boldsymbol{p}\, \delta\left(E - \sqrt{p_1^2 + p_2^2}\right) \\
&\quad \times \left(\frac{\alpha_1^2\Delta^2 p_1^2}{\alpha_1^2\Delta^2 + v_{\mathrm{F}}^2} + \frac{\alpha_2^2\Delta^2 p_2^2}{\alpha_2^2\Delta^2 + v_{\mathrm{F}}^2}\right) \left((\varepsilon_n(\boldsymbol{k}_{\mathrm{n}}) - \varepsilon_m(\boldsymbol{k}_{\mathrm{n}}))^2 g_{\mu\nu}^{nm}(\boldsymbol{k}_{\mathrm{n}}) + \mathcal{O}(E)\right) \\
&= \frac{1}{2}\left(\frac{\alpha_1^2\Delta^2}{\alpha_1^2\Delta^2 + v_{\mathrm{F}}^2} + \frac{\alpha_2^2\Delta^2}{\alpha_2^2\Delta^2 + v_{\mathrm{F}}^2}\right) \left\langle (\varepsilon_n(\boldsymbol{k}) - \varepsilon_m(\boldsymbol{k}))^2 g_{\mu\nu}^{nm}(\boldsymbol{k}) \right\rangle_{E=0,n} E + \mathcal{O}(E^2).
\end{aligned} \tag{S.81}$$

In a similar manner, the low-energy behavior of the expectation values in the leading order can be evaluated for other cases. The results are summarized in Table S.III.

| gap type | (a) flat & Dirac bands | (b) Dirac band | 3. dispersive & dispersive bands |
|---|---|---|---|
| full gap | $(E^2 - \Delta^2)^{-1}$ | const. | const. |
| point nodes | $E^{2-2/l}$ & $E^{2(l-1)}$ | $E^{2l-1}$ | $E$ |
| line nodes (no crossing) | $E^{2-1/l}$ & $E^{2(l-1)}$ | $E^{2l-1}$ | $E$ |
| line nodes (with crossing) | $-E^{2-1/l}/\ln E$ & $E^{2(2l-1)}$ | $E^{4l-1}$ | $E$ |

Supplementary Table S.III. Low-energy behaviors of the expectation values that appear in Eq. (S.56) or Eq. (S.58) in the presence of band crossing at the Fermi surface. We consider nodal structures characterized by $l = s = 1$ or 2, corresponding to linear or quadratic dispersion near the gap node. In the "flat & Dirac bands" column for nodal gaps, the left entry corresponds to the expectation value associated with the flat band, and the right entry to that associated with the Dirac band.

### III. Temperature Dependence of Quantum-Geometric Superfluid Weight

Using the density of states and the expectation values derived in Sections I and II, we now evaluate the temperature dependence of the quantum-geometric contribution to the superfluid weight $D_{\mu\nu}^{\mathrm{geom}}$ based on Eq. (S.56). We present here several representative cases.

## A. Full Gap

We begin with the analysis of a flat band without band crossing, assuming a full-gap superconducting state. In this case, the temperature derivative of the geometric contribution becomes

$$\frac{\partial \delta D_{\mu\nu}^{\text{geom}}(T)}{\partial T} \propto -\frac{1}{T}\int_0^\infty dE f'(E)\delta(E-\Delta) \propto \frac{1}{T^2}\int_0^\infty dE \frac{e^{\beta E}}{(e^{\beta E}+1)^2}\delta(E-\Delta)$$
$$= \frac{1}{T^2}\frac{e^{\beta\Delta}}{(e^{\beta\Delta}+1)^2} \simeq \frac{1}{T^2}e^{-\Delta/T}, \tag{S.82}$$

which leads to the following result for the geometric correction:

$$\delta D_{\mu\nu}^{\text{geom}}(T) \propto \int_0^T dx\, x^{-2} e^{-\Delta/x} \simeq \Delta^{-1} e^{-\Delta/T} \propto e^{-\Delta/T}. \tag{S.83}$$

Here, we have used the asymptotic expression valid in the low-temperature limit ($T \to 0$),

$$\int_0^T dx\, x^{-a} e^{-\Delta/x} = \Delta^{-a+1}\int_{\Delta/T}^\infty dy\, y^{(a-1)-1} e^{-y} = \Delta^{-a+1}\Gamma(a-1, \Delta/T) \simeq \Delta^{-1} T^{-a+2} e^{-\Delta/T}, \tag{S.84}$$

where $\Gamma(a,x)$ denotes the incomplete Gamma function.

Next, analyze the case where a flat band crosses a Dirac band. In this case, we obtain the result,

$$\frac{\partial \delta D_{\mu\nu}^{\text{geom}}(T)}{\partial T} \propto -\frac{1}{T}\int_0^{E_c} dE\Big(f'(\Delta)-f'(E)\Big)E\Theta(E-\Delta)\frac{1}{E^2-\Delta^2}$$
$$\propto \frac{1}{T^2}\lim_{\delta\to 0^+}\int_{\Delta+\delta}^{E_c} dE\left(\frac{e^{\beta\Delta}}{(e^{\beta\Delta}+1)^2}-\frac{e^{\beta E}}{(e^{\beta E}+1)^2}\right)\frac{E}{E^2-\Delta^2} \tag{S.85}$$
$$\simeq \frac{1}{T^2}\ln\left(\frac{E_c^2-\Delta^2}{2\Delta}\right)e^{-\Delta/T},$$

which yields

$$\delta D_{\mu\nu}^{\text{geom}}(T) \propto e^{-\Delta/T}. \tag{S.86}$$

Here, $E_c$ is a high-energy cutoff.

## B. Nodal Gap

In all cases except for the crossing line nodes, the temperature dependence of the quantum-geometric superfluid weight is generally of the form,

$$\frac{\partial \delta D_{\mu\nu}^{\text{geom}}(T)}{\partial T} \propto -\frac{1}{T}\int_0^\infty dE f'(E)E^a \propto \frac{1}{T^2}\int_0^\infty dE\frac{e^{\beta E}}{(e^{\beta E}+1)^2}E^a = a\Gamma(a)\eta(a)T^{a-1}, \tag{S.87}$$

where $\eta(x)$ is the Dirichlet eta function. Upon integration over temperature, this yields the leading-order low-temperature behavior,

$$\delta D_{\mu\nu}^{\text{geom}}(T) \propto T^a. \tag{S.88}$$

For the case of crossing line nodes, the energy dependence acquires a logarithmic correction, leading to

$$\frac{\partial \delta D_{\mu\nu}^{\text{geom}}(T)}{\partial T} \propto \frac{1}{T}\int_0^\infty dE f'(E)E^a \ln E \propto -\frac{1}{T^2}\int_0^\infty dE\frac{e^{\beta E}}{(e^{\beta E}+1)^2}E^a \ln E \simeq -a\Gamma(a)\eta(a)T^{a-1}\ln T, \tag{S.89}$$

which yields

$$\delta D_{\mu\nu}^{\text{geom}}(T) \propto -T^a \ln T. \tag{S.90}$$

Only the leading-order contributions have been retained in the low-temperature limit $T \to 0$.

In models in the presence of band crossing, such as models (b) and (c), we encounter a different temperature dependence due to the degeneracy at the Fermi level. In such cases, we obtain

$$\frac{\partial \delta D_{\mu\nu}^{\text{geom}}(T)}{\partial T} \propto \frac{1}{T} \int_0^\infty dE f''(E) E^a \propto \frac{1}{T^3} \int_0^\infty dE \frac{e^{\beta E}}{(e^{\beta E}+1)^2} \tanh\left(\frac{\beta E}{2}\right) E^a \qquad \text{(S.91)}$$
$$= a(a-1)\Gamma(a-1)\eta(a-1)T^{a-2},$$

leading to the power law,

$$\delta D_{\mu\nu}^{\text{geom}}(T) \propto T^{a-1}. \qquad \text{(S.92)}$$

By performing these calculations, we obtain low-temperature behaviors of the quantum-geometric superfluid weight for various gap structures and band structures.

### C. General Results

Finally, we summarize the low-temperature behavior of the conventional and quantum-geometric contributions to the superfluid weight for various nodal structures of the superconducting gap. Table S.IV presents the results for systems without band crossing on the Fermi surface, while Table S.V shows the corresponding results for systems with band crossing. In the main text, we focus on the physically relevant cases where $l = s$, corresponding to nodal structures with linear or quadratic dispersion.

| gap structure | $\delta D_{\mu\nu}^{\text{conv}}$ | $\delta D_{\mu\nu}^{\text{geom}}$ (flat band) | $\delta D_{\mu\nu}^{\text{geom}}$ (dispersive band) |
|---|---|---|---|
| full gap | $T^{-1/2} e^{-\Delta/T}$ | $e^{-\Delta/T}$ | $T^{1/2} e^{-\Delta/T}$ |
| point node | $T^{1/l+1/s}$ | $T^{1/l+1/s+1}$ | $T^{1/l+1/s+2}$ |
| line node (w/o crossing) | $T^{1/l}$ | $T^{1/l+1}$ | $T^{1/l+2}$ |
| line node (with crossing) | $-T^{1/l} \ln T$ | $-T^{1/l+1} \ln T$ | $-T^{1/l+2} \ln T$ |

Supplementary Table S.IV. Temperature dependence of the conventional and quantum-geometric contributions to the superfluid weight in the absence of band crossing. For point nodes, $l = 1$ and $s = 1$ correspond to linear point nodes, whereas $l = 2$ and $s = 2$ correspond to quadratic point nodes. Similarly, for line nodes, $l = 1$ represents linear line nodes, while $l = 2$ represents quadratic line nodes.

| gap structure | (a) flat & Dirac bands | (b) Dirac band | (c) dispersive & dispersive bands |
|---|---|---|---|
| full gap | $e^{-\Delta/T}$ | $e^{-\Delta/T}$ | $T^{-1/2} e^{-\Delta/T}$ |
| point node | $T$ | $T^{2l-1}$ | $T^{2/l}$ |
| line node (w/o crossing) | $T$ | $T^{2l-1}$ | $T^{1/l}$ |
| line node (with crossing) | $T$ | $T^{4l-1}$ | $-T^{1/l} \ln T$ |

Supplementary Table S.V. Temperature dependence of the quantum-geometric contribution to the superfluid weight in the presence of band crossing. The parameter $l$ takes a value of either 1 or 2, corresponding to linear and quadratic nodes, respectively.


\fi

\end{document}